\documentclass[sigconf]{acmart}
\usepackage{soul}
\usepackage{afterpage}
\usepackage{xspace}
\usepackage{longtable}
\usepackage{fontawesome5}
\usepackage{graphicx}
\usepackage{subcaption}
\usepackage{tabularx}
\usepackage{multirow}
\usepackage{multicol}
\usepackage{enumitem}
\usepackage{longtable}
\usepackage{caption}
\definecolor{pumpkin}{RGB}{211, 84, 0}

\newcommand{\zzrevision}[1]{\textcolor{black}{#1}}

\AtBeginDocument{%
  \providecommand\BibTeX{{%
    \normalfont B\kern-0.5em{\scshape i\kern-0.25em b}\kern-0.8em\TeX}}}

\copyrightyear{2025}
\acmYear{2025}
\setcopyright{cc}
\setcctype{by}
\acmConference[CHI '25]{CHI Conference on Human Factors in Computing Systems}{April 26-May 1, 2025}{Yokohama, Japan}
\acmBooktitle{CHI Conference on Human Factors in Computing Systems (CHI '25), April 26-May 1, 2025, Yokohama, Japan}\acmDOI{10.1145/3706598.3713289}
\acmISBN{979-8-4007-1394-1/25/04}

\begin{document}

\title[LADICA: A Large Shared Display for Cognitive Assistance in Co-Located Team Collaboration]{LADICA: A Large Shared Display Interface for Generative AI Cognitive Assistance in Co-Located Team Collaboration}


\author{Zheng Zhang}
\affiliation{%
  \institution{University of Notre Dame}
  \city{Notre Dame}
  \state{IN}
  \country{USA}
}

\author{Weirui Peng}
\affiliation{%
  \institution{Columbia University}
  \city{New York}
  \state{NY}
  \country{USA}
}

\author{Xinyue Chen}
\affiliation{%
  \institution{University of Michigan}
  \city{Ann Arbor}
  \state{MI}
  \country{USA}
}

\author{Luke Cao}
\affiliation{%
  \institution{University of Notre Dame}
  \city{Notre Dame}
  \state{IN}
  \country{USA}
}

\author{Toby Jia-Jun Li}
\affiliation{%
  \institution{University of Notre Dame}
  \city{Notre Dame}
  \state{IN}
  \country{USA}
}

\renewcommand{\shortauthors}{Zheng Zhang, Weirui Peng, Xinyue Chen, Luke Cao, Toby Jia-Jun Li}

\begin{abstract}
Large shared displays, such as digital whiteboards, are useful for supporting co-located team collaborations by helping members perform cognitive tasks such as brainstorming, organizing ideas, and making comparisons. While recent advancement in Large Language Models (LLMs) has catalyzed AI support for these displays, most existing systems either only offer limited capabilities or diminish human control, neglecting the potential benefits of natural group dynamics. Our formative study identified cognitive challenges teams encounter, such as diverse ideation, knowledge sharing, mutual awareness, idea organization, and synchronization of live discussions with the external workspace. In response, we introduce LADICA, a large shared display interface that helps collaborative teams brainstorm, organize, and analyze ideas through multiple analytical lenses, while fostering mutual awareness of ideas and concepts. Furthermore, LADICA facilitates the real-time extraction of key information from verbal discussions and identifies relevant entities. A lab study confirmed LADICA's usability and usefulness.

\end{abstract}



\begin{teaserfigure}
  \includegraphics[width=\textwidth]{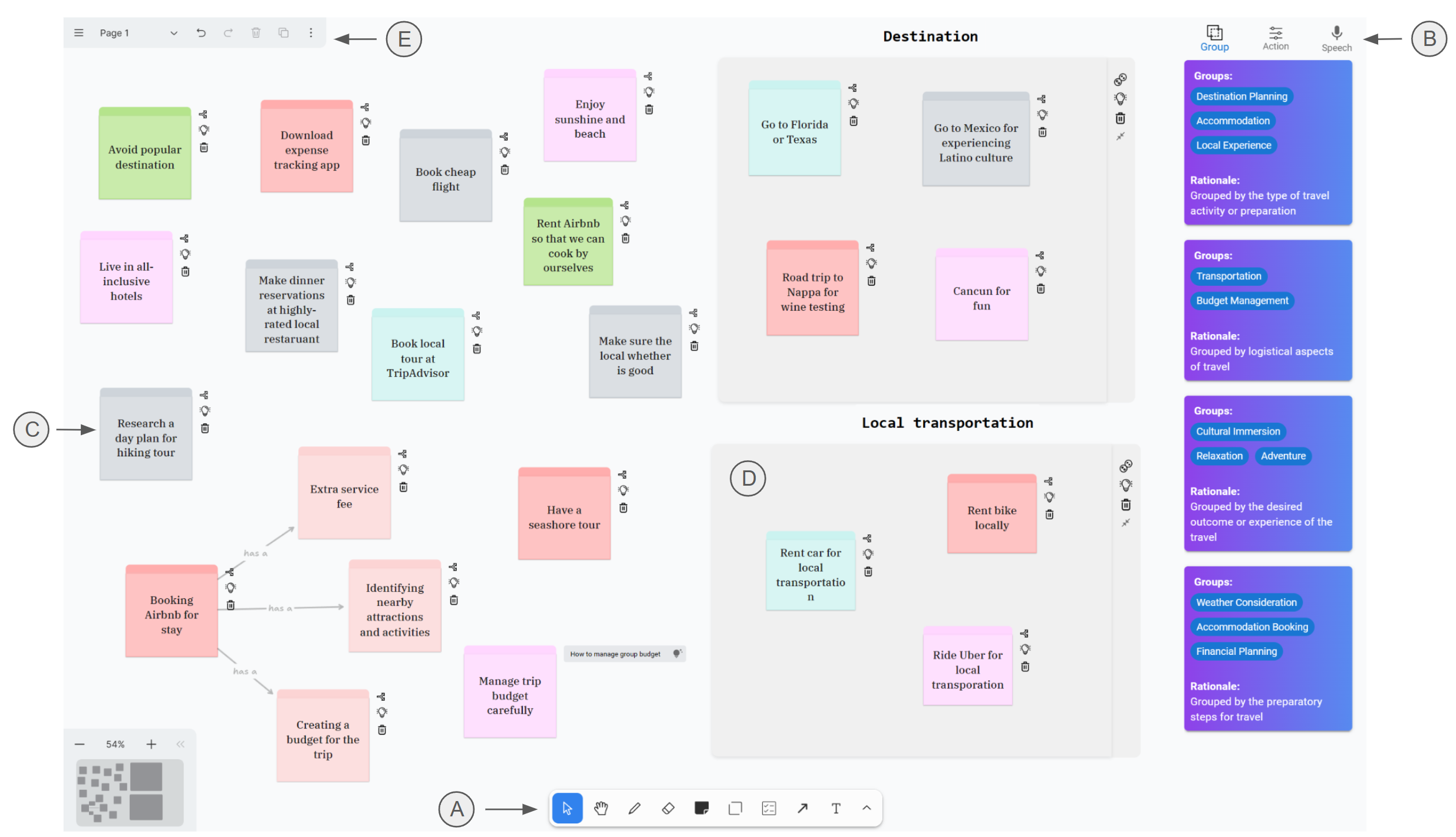}
  \caption{The main interface of LADICA. Through the toolbar (A), teams can generate various objects on a shared display, including idea notes (C) and topic groups (D). LADICA enables team members to simultaneously edit content via personal devices or on the shared touch display, with their identities distinguished by color. Teams can access various AI cognitive assistance features for idea notes and topic groups. In addition, they can use global functions like global affinity-based grouping, relation hint generation, and discussion-based information extraction and retrieval from the workspace menu (B). Teams can navigate between different grouping views using the drop-down menu (E). }
  \label{fig:teaser}
\end{teaserfigure}
\maketitle

\section{INTRODUCTION}





Co-located collaboration, where team members work together in the same location and at the same time, often leads to improved productivity for tasks that benefit from frequent communication and joint efforts, such as group brainstorming, knowledge building, and planning~\cite{westendorf2017understanding, andrews2010space, olson2000distance, wang2006providing, hmelo2008facilitating}. To support group work, large shared displays, such as digital whiteboards, are frequently used to offload ideas, create a repository of shared memories, and organize information. Compared to traditional non-digital whiteboards, digital displays provide a scalable workspace, flexible editing capabilities, and on-screen tools such as sticky notes, highlights, and freestyle drawing to facilitate team creation.



While digital displays were initially used mainly as tools for preserving and presenting information, prior work has explored extending them to provide intelligent support for group cognitive tasks. Existing systems have adopted basic machine learning (ML) approaches to provide cognitive support, such as semantic grouping, keyword extraction, and inspirational stimuli matching~\cite{shi2017ideawall, andolina2015inspirationwall, koch2020imagesense}. However, these features are not attuned to the dynamics of human cognition and collaboration. This results in superficial support that fails to deeply engage with the team's cognitive processes.

Recent advancements in large language models (LLMs) have showcased their capabilities in knowledge understanding, generation, and reasoning. Leveraging these developments, various AI-enhanced team collaboration tools have incorporated LLMs to support group activities. Platforms such as Miro\cite{miro}, Microsoft Teams\cite{microsoft:teams}, and Lucidspark~\cite{lucidspark} now feature AI-powered tools for idea generation, grouping, and summarization. However, the design of these tools primarily focuses on direct capabilities of generative AI models, rather than deeply understanding how technology could improve traditional human practices and the complex cognitive processes in teamwork.


On the other hand, some AI tools have been introduced to automate thinking processes on behalf of human teams~\cite{liu2018consensus, chen2023meetscript}. While these tools might boost productivity in team collaboration, they can lead to an over-reliance on AI in the thinking and discussion process. This can be counterproductive, especially in co-located collaboration, where the essence of collaboration lies in human interaction and engagement~\cite{olson2000distance, ez2023group}. To strike a balance, AI-enhanced collaboration tools should aim to supplement and facilitate human cognitive activities, not replace them. The goal is for AI to provide timely and relevant support while preserving space for human creativity, critical thinking, and social interaction \cite{wang2020human, piper2011augmenting}.

In this paper, we aim to design and develop an intelligent large shared display interface that incorporates LLM assistance into human-led co-located group collaboration \zzrevision{for brainstorming and sensemaking activities}. The primary goal is to provide dynamic cognitive support that enhances and supports beneficial practices in human-human  interactions. In a formative study aimed at understanding how teams use whiteboards during in-person collaboration, we found that coworkers frequently rely on a shared whiteboard for brainstorming, grouping analysis and grounding their verbal discussions. This study pinpointed several key challenges participants face in such collaborations, as outlined Section \ref{sec:formative_findings}. Through conducting scenario-based design workshops, we gathered insights into users' concerns about the AI dominance and the potential over-reliance on AI in group collaborations. From these insights, we identified four specific design goals: (1) facilitate the cognitive process of task breakdown and diverse ideation; (2) enhance collaboration by fostering mutual awareness; (3) offer support for comprehensive analysis from various perspectives and dimensions; (4) encourage synchronization among team members of discussion entities and workspace activities.



\begin{figure*}
    \centering
    \includegraphics[width=\linewidth]{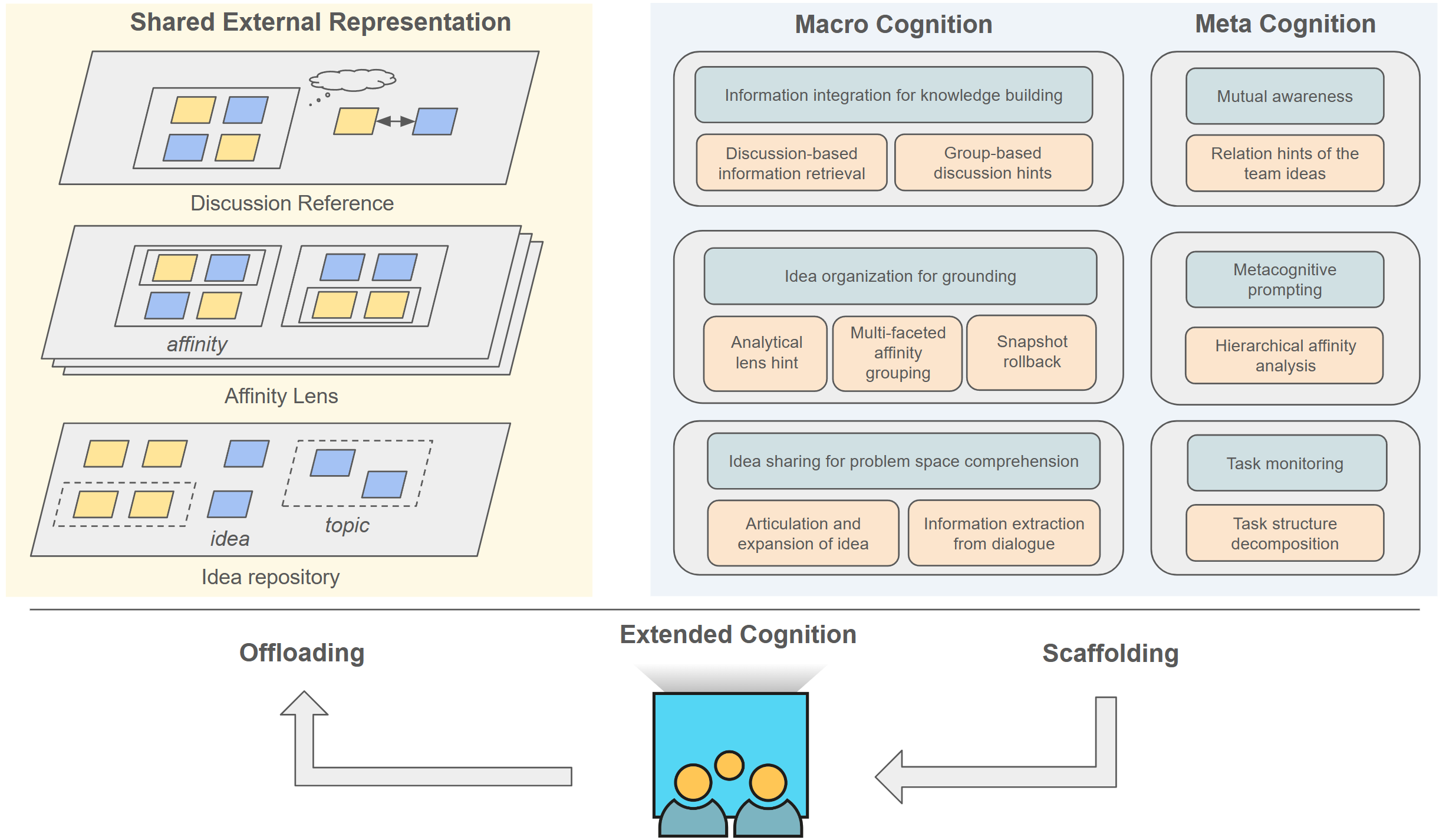}
    \caption{Conceptual framework of LADICA. Teams build external cognition by offloading their ideas and information shared among the team as external representations on the large shared display (left), and these external representations act as scaffolds to support teams' cognitive process, enabling them to discuss and coordinate more effectively (right). During the offloading stage, teams create three layers of representation, idea repository, affinity lens, and discussion reference. Each representation layer scaffolds meta- and marco-cognitive processes in corresponding team activities (group ideation, analysis and discussion-based knowledge-building) respectively. }
    \label{fig:conceptual_framework}
\end{figure*}

Based on those findings, we introduce LADICA\footnote{LADICA stands for \textbf{LA}rge \textbf{DI}splay with \textbf{C}ognitive \textbf{A}ssistance}, an AI-enhanced large shared display interface designed to support co-located team's meta- and macro-cognitive processes of ideation, grouping analysis, and discussion. LADICA augments the traditional memory offloading function of a shared display by providing cognitive scaffolding throughout the co-located collaboration process. Drawing inspiration from Fiore's cognitive framework of collaborative knowledge building~\cite{fiore2018data}, LADICA adopts a novel approach to support team ideation, idea organization and analysis, and promote knowledge-building discussions in three layers of representation: \textit{idea repository}, \textit{affinity lens}, and \textit{discussion reference}. Each layer serves as a unique external memory representation and offers tailored cognitive assistance for co-located ideation, analysis, and discussion. Through this design, LADICA encourages beneficial practices such as creating a shared understanding of the team's task structure, expanding individual ideas, analyzing group ideas from various analytical perspectives, creatng mutual awareness and fostering dialogue among team members. This three-layer approach is illustrated in Figure \ref{fig:conceptual_framework} and elaborated in Section \ref{sec:layers}. \looseness=-1



We conducted a lab user study with 14 participants to evaluate LADICA's usability and usefulness. The results of this study indicate that users were able to successfully leverage LADICA to enhance their co-located group collaboration. Participants provided valuable insights into the dynamic roles of LADICA in the co-located collaboration process and expressed appreciation for the system's support in promoting divergent thinking, inclusive participation, and the development of a shared mental model. They also appreciated how LADICA helped them align internal and external memory during discussions. Furthermore, participants offered feedback for further improvements to the system.\looseness=-1



In summary, this paper makes the following contributions:

\begin{itemize}
    \item A formative study aimed to understand the practices and cognitive challenges encountered by co-located teams, as well as their desires and concerns regarding the use of AI to enhance co-located group collaboration.
    \item The development of LADICA, a large shared display interface that utilizes LLMs to provide multifaceted cognitive assistance using a three-layer approach in co-located group collaboration \zzrevision{for brainstorming and sensemaking activities}.
    \item A user study with 14 participants, which validated the usability and usefulness of LADICA in supporting co-located group collaboration.
\end{itemize}

\section{BACKGROUND AND RELATED WORK}

\subsection{Co-Located Group Collaboration}


Co-located group collaboration, according to the groupware taxonomy that classifies by time and place~\cite{ellis1991groupware}, describes situations in which individuals work together on a task simultaneously and at the same physical location. One of its key properties is that group members have mutual access to shared artifacts in their workspace. 
Olson et al.~\cite{olson2000distance} argued that co-located collaboration provides multiple favorable perceptual and cognitive characteristics:
\begin{itemize}
    \item \textbf{Implicit cues}: working in the same location allows coworkers to discern a variety of nonverbal cues going on in the periphery, such as body posture, gesture, voice and facial expression. Noticing those nuances and natural operations of coworkers' attention provides important contextual information unavailable in remote settings.
    \item \textbf{Rapid feedback}: the close proximity enables coworkers to provide rapid feedback. This facilitates quick opportunities for collaboration and allows for the timely resolution of misunderstandings and disagreements as soon as they arise.
    \item \textbf{Coreference}: In co-located interactions, participants can use their gaze or gesture to indicate the referent of deictic terms such as "this" or "that". This direct gaze alignment helps to disambiguate referential expressions and ensures that all participants are focused on the same entity. Additionally, the shared local context enables individuals to spatially reference objects, enhancing mutual understanding of each other's thoughts.
\end{itemize}


Meanwhile, the ``same time, same location'' presence of co-located collaboration introduces several
social-motivational hurdles that hurt group productivity in co-located collaboration. 
First, the physical presence of others makes people more conscious of how they are perceived in others' eyes, known as evaluation apprehension~\cite{nijstad2006group}. Individuals who are more sensitive to being judged often feel anxious and hesitant to share their ideas and fear negative feedback. Another source of productivity loss is production blocking~\cite{nijstad2003production}. Collaborators often need to express their ideas spontaneously or immediately after the ideas come to mind. However, in co-located setting, methods such as taking turns to speak can lead people to forget or overthink their ideas, resulting in them choosing not to share with the group.

Within this framework, our study excludes loose coupling scenarios, such as coordination through asynchronous communication tools.  Instead,  we focus on co-located collaboration characterized by tight coupling \cite{stuckel2008effects} and high synchronicity \cite{linebarger2005benefits}, where immediate and face-to-face interaction is essential for task completion. These scenarios involve close interaction among participants, where rapid feedback, implicit cues, and co-reference play crucial rules in enhancing effective collaboration.


\subsection{Team Cognition}
\label{sec:cognition_theory}

Team cognition is central to team effectiveness. Warner et al.~\cite{warner2005cognitive} proposed a cognitive model of the team cognition consisting process at two levels: Meta-cognition and Macro-cognition.

\textbf{Meta-cognition}, defined as "cognition about cognition", involves team members thinking about how the team processes information, works on problems, and feels about the team process \cite{thompson2012metacognition}. Prior study shows that teams with strong metacognitive skills can better monitor their progress, adapt to changes, and leverage diverse perspectives, leading to improved outcomes \cite{nonose2014effects}.   One key metacognitive activity is \textit{task monitoring} - Teams should regularly monitor their progress towards their goals, assessing to which they clearly understand their task approach \cite{kim2018effect}. Besides, teams can benefit from prompts or questions that encourage them to reflect on their thinking and decision-making processes, which is called as \textit{metacognitive prompting}~\cite{wiltshire2014training}. Additionally, \textit{meta-cognitive awareness} helps teams recognize their cognitive strengths and weaknesses, allowing for more informed and adaptive problem-solving \cite{martirosov2021team}. 


On the other hand, \textbf{macro-cognition} is the process by which teams create new knowledge through internalized and externalized mental processes \cite{letsky2007macrocognition}. Fiore et.al decomposes macro-cognition into three layers \cite{fiore2018data}. The first layer is the "data" layer, where the initial information is shared in the conversation or in a shared external space without any mutual understanding yet established \cite{fiore2018data}. This process involves receiving of parts, and solution option generation, where ideas are proposed and expanded, and the team knowledge starts to be constructed \cite{fiore2010towards}. The second layer comes to the "information" layer, in which the group gives meaning to data through relational connection and interpretation \cite{fiore2018data}. In the "knowledge" layer, team members then discuss to structure the information into a coherent format that can be easily understood \cite{fiore2018data}, so as called as \textit{"assimilation or accretion of knowledge"} \cite{fiore2010towards}.

Prior works have mostly focused on facilitating offloading and scaffolding process of team cognition. The \textbf{offloading} process primarily serves as a memory aid, replacing the cognitive process where people hold items in their working memory or retrieve them from long-term memory \cite{clark2008supersizing}. To improve this process, meeting visualization systems~\cite{chandrasegaran2019talktraces, chen2023meetscript} were designed to offload verbal communication with visuals in order to reduce cognitive efforts in synchronous communication. Similarly, commercial tools, such as Miro, LucidChart, were commonly used in team collaboration as a way for people to offload working memory and visualize ideas \cite{johnson2022miro}.  \textbf{Scaffolding} is another critical category of external cognition, which entails externalizations of cognition that directly support team-level processes \cite{nussbaum2009technology}. Technological scaffolds can enhance team collaboration by facilitating the exchange, comparison, and evaluation of ideas and information, and by boosting social-cognitive interactions that improve conversation and communication \cite{fiore2016technology}. For instance, IdeaExpander was designed to support group brainstorming by presenting pictorial stimuli of verbal conversation \cite{wang2010idea}.   

Our work builds upon the aforementioned theories of team cognition to support teams extending their cognition through offloading their ideas and thinking process using a large shared display, and designed a set of functions to scaffold both meta- and macro-cognitive processes of collocated team collaboration.

 \subsection{Shared Displays for Supporting Co-Located Interaction}


Multi-touch, large-size displays are frequently used as a means to facilitate interaction in co-located settings. A large digital display offers enhanced flexibility for social interaction and spatial organization within a scalable space~\cite{olsson2020technologies}. As a shared workspace, those displays could visualize and externalize people's working memory to offload group thinking~\cite{luck2013visual, risko2016cognitive, fiore2016technology}, and enable a range of gestural interactions that bolster cooperation and engagement~\cite{hascoet2008throwing}.

The prior works in the field of large shared display have largely focused on \textit{enabling, enhancing, and encouraging} co-located interaction on shared display. The primary focus of enabling research is on innovating interaction techniques suitable for co-located interactions. This encompasses a variety of input methods, including projection-based~\cite{shoemaker2007shadow, hereld2000introduction}, touch-based~\cite{hilliges2009interactions, vogel2004interactive, kim2010multi}, and gaze-based \cite{zhai1999manual, muller2009reflectivesigns} techniques. Besides enabling technology, prior works also studied different visualization techniques to enhance group brainstorming and sensemaking~\cite{shi2017ideawall, langner2018multiple}.  Lastly, some works focused on designing interfaces that encourage interactions among individuals, such as gamify collaboration~\cite{lanzilotti2015pupils} and content sharing~\cite{wallace2011investigating}.

With the recent advance in LLM, some commercial collaboration tools like Miro~\cite{miro}, Microsoft Team~\cite{microsoft:teams} and Lucidspark~\cite{lucidspark} have incorporated LLM-empowered features to facilitate team collaboration, such as mind map generation, idea expansion and grouping. Despite its benefits, the cognitive support for co-located collaboration remains limited. Co-located settings prioritize frequent exchanges of thoughts and collaborative work in a shared space. Simply presenting optimal ideas or group results does not adequately support the dynamic cognitive processes, group thinking, and communication essential in these environments.

Our work builds upon previous efforts to leverage LLM for enhancing team collaboration. Specifically, we aim to integrate LLM to offer meta- and marco-cognitive support, thereby promoting beneficial practices in human collaboration and communication that are crucial for successful co-located teamwork.

\section{Formative Study}

We conducted a formative study comprising a focus group and a scenario-based design workshop. This study aimed to explore the behaviors, cognitive challenges, and needs of teams during face-to-face group collaborations. Additionally, we sought to gather insights into their opinions, preferences, and concerns about how AI could be leveraged to enhancing co-located group collaboration.

Specifically, the study focused on uncovering: (1) the main uses of a shared whiteboard in co-located collaboration; (2) the challenges faced by co-located team members; (3) the desired features of a shared digital whiteboard to support co-located teamwork; (4) users' concerns regarding AI intervention during team collaboration.

\subsection{Process}


\subsubsection{Participants}

We recruited eight participants (five males, three females) with experience in in-person team collaboration from a university's mailing list. Each participant was compensated with a \$15 USD gift card. Our participants consisted of four undergraduate and four graduate students, with majors in computer science, business, applied math, and social science. They had participated in various forms of in-person teamwork, including planning club events, discussing course projects, organizing trips, and planning software development. \zzrevision{Seven of them have used AI products (e.g. ChatGPT, Gemini) in the past. One participant had little to no knowledge of AI, four had a basic understanding of AI concepts, and three had experience developing AI systems or possessed a strong understanding of AI.} Besides, they all had previously used a shared physical whiteboard for collaborative teamwork and had experience with digital tools such as Excel, Notion, Miro, and Microsoft OneNote for group work.


\subsubsection{Procedure} 

Our study involved two in-person sessions, each with four participants, lasting one hour. \zzrevision{Two authors attended each session to facilitate the study and take notes.} These sessions were divided into a 30-minute focus group and a 30-minute scenario-based design workshop\footnote{The study protocol has been reviewed and approved by the IRB at our institution.}. In focus groups, participants discussed their past experiences, challenges, and unmet needs when working with a co-located team. Additionally, we also asked them to discuss their concerns about AI intervention during team collaboration. 

Following the focus group, we presented a scenario of using an AI-enhanced whiteboard to discuss and plan a spring break trip with friends. Participants were invited to illustrate their ideal AI features on paper storyboards or write down their ideas on sticky notes. The study sessions were recorded \zzrevision{using an external video camera}, then transcribed and analyzed using the reflexive thematic analysis method~\cite{braun2012thematic}.


\subsection{Findings}
\label{sec:formative_findings}

\subsubsection{Whiteboard is used for supporting ideation, analysis and conversation}
\label{sec:finding_of_cognitive_process}

In our formative study, we found that in-person teams primarily used a shared whiteboard for three key purposes: visually expressing individual ideas, collectively understanding and analyzing tasks, and providing a visual anchor for discussions. \zzrevision{Through the focus group discussion, participants have agreed that} visualization helps align members' mental models, aiding in \textit{``avoiding knowledge gaps in verbal discussions}'' (P3) and \textit{``reducing the cognitive load of remembering shared information}'' (P6). Additionally, as teams often focus on ideation and planning stages in meetings (P3, P4, P5, P7, P8), the whiteboard supports both bottom-up and top-down approaches to developing goals. \zzrevision{For example, during design workshops, P7 brainstormed features of meaningfully organizing diverse ideas in open-ended ideation, while P4 sketched a more structured approach to create a top-down plan and then delegate tasks
.} Whiteboards also help teams base discussions on previously visualized information, allowing them to ``\textit{refer back to clarify which ideas are being discussed and identify new ideas based on the existing ones}'' (P2). Based on these insights, we are developing a digital whiteboard system designed to enhance cognitive processes during group ideation, analysis, and conversation, and to support flexible collaboration styles.

\subsubsection{\textit{Cognitive and social motivational challenges hinder ideation and knowledge sharing process.}}

Consistent with prior research~\cite{ez2023group, dijkstra2008social, drapeau2014sparking}, our participants highlighted cognitive and socially motivational challenges during in-person collaboration. \zzrevision{The focus groups showed that} individuals often struggle with divergent and critical thinking, finding it difficult to \textit{``quickly generate a wide range of ideas or comment on others' ideas}'' (P6) and to ``\textit{articulate thoughts relevantly and clearly}'' (P4). Socially, the presence of team members can create pressures that reduce effectiveness, such as conformity pressure leading to narrow idea selection (P3, P5) and fear of judgment inhibiting idea sharing (P1, P3, P7). Additionally, teams may find it challenging to understand task structure and prioritize topics and subtasks during brainstorming (P3, P8). To address these issues, participants suggested integrating AI features to inspire and refine ideas, facilitate expression, and clarify discussion points and task structures, aiming to enhance both the creative and organizational facets of team collaboration.

\subsubsection{\textit{Maintaining mutual awareness during parallel team work is crucial yet difficult}} 

Mutual awareness is crucial for managing interdependent tasks and overall progress in teams~\cite{liu2016shared, tang2006collaborative}. \zzrevision{The focus groups} identified challenges in maintaining this awareness during parallel tasks within a shared workspace. For instance, P5 noted, \textit{``when working on different parts of a task on a whiteboard, it's challenging to remain mindful of how my work relates to and impacts others and vice versa. This awareness is crucial for us to coordinate interdependent and parallel tasks}.'' P7 added,\textit{``we tend to miss collaboration opportunities when people focus on different parts without realizing the connection among each other's ongoing work}.'' To mitigate these issues, participants recommended integrating AI to highlight connections between individual contributions, enhancing understanding of how ideas interrelate and revealing opportunities for dialogue.


\subsubsection{The need for varied analytical perspectives in organizing and comparing information}

Participants noted that whiteboards are crucial for organizing group ideas into themes, fostering a bottom-up understanding, stimulating informal discussions, and identifying goal-related gaps. They also expressed a need for diverse analytical perspectives to organize information and track updates, essential for revealing common themes and patterns in their proposals (P6). However, the current digital whiteboard tools' limited support for flexible information grouping and tracking perspective evolution restricts experimenting with various organizational methods and gaining deeper insights. \zzrevision{In the design workshops,} some participants (P3, P5) sketched features for saving and reloading different workspace snapshots, facilitating experimentation with various groupings and explorations. In response, during design workshops, participants envisioned scenarios where AI could suggest different categorization strategies for team-generated ideas and automatically group them, enhancing the exploration of various analytical perspectives.

\subsubsection{Challenges in Synchronizing and Anchoring Discussion on Shared Content}

In co-located group collaboration, discussions are typically informal, active, and fluid, which enhances team cohesion and engagement. However, participants noted challenges in maintaining synchronization between the discussion and the content on the whiteboard (P2, P5, P8). For example, P3 stated, ``\textit{It's easy for interesting points to be lost when everyone is engaged in the conversation and no one is documenting them on the whiteboard}.'' Moreover, when the whiteboard is overloaded with information, it becomes difficult for teams to quickly locate and concentrate on details relevant to the ongoing conversation (P3, P4, P7), which is crucial for anchoring discussions and refreshing shared understanding (P3, P4). To address these issues, participants recommended developing features that could automatically extract key points from discussions and emphasize elements on the whiteboard pertinent to the current conversation.

\subsubsection{Concerns regarding AI involvement in group collaboration}

While participants appreciated the AI features designed to enhance team collaboration, they also voiced several concerns about AI involvement. There was worry that AI might dominate the natural dynamics of human collaboration, with fears that AI-driven interactions could overshadow human-to-human interactions (P1, P3, P4, P6, P7). As P5 mentioned, ``\textit{I have little worry about AI taking over control of discussions, though I feel like reasonable humans still know what they want from the discussion}.'' Additionally, there were concerns that relying on AI for idea generation and planning could reduce critical thinking and creativity among team members (P2, P4, P5, P8). Privacy issues were also raised, with a preference expressed for AI assistance to be optional and only activated upon explicit request (P1, P4, P7).

\subsection{Design Goals}

Based on our formative study findings and theories on the cognitive processes of team collaboration introduced in Section \ref{sec:cognition_theory}, we have identified four design goals for an LLM-enhanced large shared display interface to facilitate co-located team collaboration:

\begin{itemize}
    \item \zzrevision{\textbf{DG1:}\textbf{ Enhancing dynamic ideation and knowledge sharing within groups}. Social and motivational barriers, such as fear of judgment and social loafing, can prevent team members from sharing ideas freely, affecting inclusivity and interaction quality. Cognitive theories and our formative study highlight the need for metacognitive support to help teams understand complex tasks. Team members also need help articulating ideas, refining them during discussions, and recognizing overlooked ideas. To address these challenges, our proposed system provides cognitive assistance to support ideation and promote knowledge sharing.}
    \item \zzrevision{\textbf{DG2:} \textbf{Streamlining bidirectional discussion and work space synchronization}. Maintaining mutual awareness is crucial for identifying collaborative opportunities, especially when team members work on different tasks simultaneously. Our formative study shows that this awareness can be difficult to achieve due to cognitive overload. While participants are open to AI assistance, they worry that too much AI involvement could disrupt natural team dynamics and reduce direct interaction. Therefore, our system is designed to help teams see how their contributions are connected, without dominating the collaboration process. It avoids automatically completing tasks that require group discussion or overly controlling the discussion flow or sequence.}
    \item \zzrevision{\textbf{DG3:} \textbf{Enabling diverse and comprehensive perspectives in analyzing shared content.} Our formative study showed that teams need to analyze and group ideas from different perspectives to uncover themes and connections, but this process can be challenging. Teams also need to track how these patterns evolve during collaboration and have the ability to save and reload workspace snapshots to experiment freely. To meet these needs, our system should support grouping analysis from multiple perspectives, helping teams in both the creation and discussion phases.}
    \item \zzrevision{\textbf{DG4:} \textbf{Fostering collaborative cues without overriding human interaction.} Co-located group collaboration often involves a lot of informal discussion. Our formative study found that synchronizing these discussions with the shared workspace can be cognitively challenging. To address this, our system identifies key points from discussions and highlights relevant information already in the workspace. This helps bridge the gap between verbal exchanges and visual representations, ensuring that important discussion elements are effectively integrated.}

\end{itemize}
\section{LADICA system}
\label{sec:system}

Drawing on the findings from the formative study, the design goals, and the cognitive theories outlined in Section \ref{sec:cognition_theory}, we developed LADICA, an AI-enhanced large shared display interface specifically tailored to support the meta- and macro-cognitive processes of ideation, grouping analysis, and discussion for co-located teams (Figure \ref{fig:teaser}). LADICA enables an automatic synchronous workspace between a shared display and personal devices, allowing co-located users to edit and manipulate simultaneously via their devices, such as tablets, phones, and laptops, or directly on the shared touch display. Idea notes created from different personal devices are displayed in distinct colors to differentiate user identities.

The remainder of this section is organized as follows: Section \ref{sec:layers} details the three representational layers of LADICA that offer cognitive support for various collaborative processes; Section \ref{sec:example} presents an example scenario demonstrating the use of LADICA in co-located collaboration; Section \ref{sec:key_features} describes the key features of LADICA; and Section \ref{sec:implementation} provides information on the implementation details.
  





\subsection{The Three-Layer Structure for Cognitive Assistance}
\label{sec:layers}

LADICA is based on a conceptual framework structured into three layers (Figure \ref{fig:conceptual_framework}), each targeting specific meta-cognitive and macro-cognitive processes in team collaboration. The framework emphasizes the role of a shared display as external memory during collaborative tasks, grounded in Fiore's cognitive framework of collaborative knowledge-building~\cite{fiore2018data}, as discussed in Section \ref{sec:cognition_theory}. We align their \textit{data-information-knowledge} process with stages of co-located ideation, analysis, and grounded discussion identified in our formative study (Section \ref{sec:finding_of_cognitive_process}). These stages facilitate cognitive processes of sharing ideas to understand the problem space (data), organizing ideas to build common ground (information), and generating new insights through discussions (knowledge).


The first layer, called the \textit{idea repository}, focuses on the initial stage of  \textit{group ideation}, where team members contribute their individual ideas, knowledge, and opinions regarding the team's goals. \zzrevision{To enhance dynamic ideation and knowledge sharing (DG1)}, this layer supports the team by organizing these individual contributions in a way that everyone can see and build upon, encouraging the expansion of ideas and a deeper understanding of the task at hand. Additionally, it helps the team decompose the overall goal into smaller, manageable tasks, assisting the team's \zzrevision{meta-cognition of understanding and monitoring} the group task structure. \zzrevision{Besides, this layer helps synchronize and offload verbal ideas to the shared workspace during the ideation phase 
(DG2).}

The second layer, the \textit{affinity lens}, aims to facilitate the \textit{organization} of shared ideas.
This layer focuses on assisting team's cognitive process of comparing and grouping ideas from various perspectives to identify structured themes within their ideas \zzrevision{(DG3)}. At the meta-cognitive level, this layer offers hierarchical affinity analysis, providing prompts that encourage hierarchical thinking. At the macro-cognitive level, this layer assists in the exploratory grouping process by offering hints for different analytical perspectives, automating multi-faceted affinity grouping, and allowing for the rollback to previously saved snapshots.

The final layer, known as the\textit{ discussion reference}, is designed to foster group discussions by highlighting the connections between different ideas. To create discussion opportunity among teams, it displays hints \zzrevision{(DG4)} about the relationships among team ideas to support the creation and maintenance of mutual awareness at the meta-cognitive level. For macro-cognitive assistance in group discussions, LADICA facilitates the connection between past and ongoing discussions by enabling the retrieval of relevant ideas on-screen during live discussions, \zzrevision{synchronizing the conversation with the content in the shared workspace (DG2)}. Furthermore, to promote collective thinking, it provides group-based discussion hints as teams focus on particular thematic groups.

\subsection{Example Scenario}
\label{sec:example}


In this section, we present a scenario demonstrating how a team uses LADICA to plan a Spring break trip using a bottom-up approach. Note that LADICA can also facilitate a top-down approach to collaboration, where teams begin by decomposing their overall goals with the help of the system (Section \ref{sec:idea_layer}).

Imagine a team consisting of Alice, Bryan, Chris, and Daniel using LADICA for trip planning. They start by activating LADICA and brainstorming for their spring break trip, recording their discussion by pressing the "Start Recording" button (Figure \ref{fig:dialogue_feature}). Each member inputs five initial ideas onto the shared whiteboard. Alice enhances his brainstorming by interacting with the \includegraphics[height=1em]{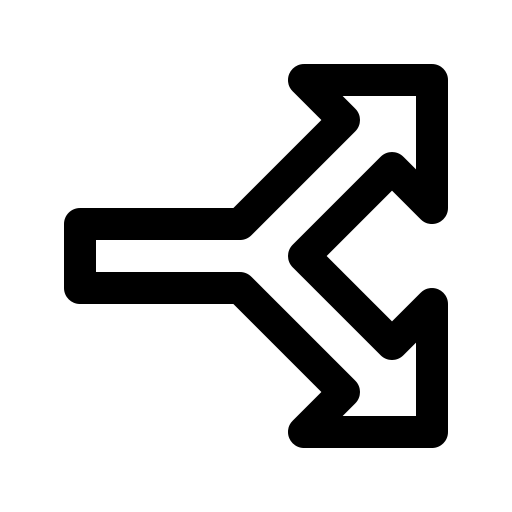} button next to his idea "booking Airbnb for stay," prompting LADICA to display a list of logical relation types (Figure \ref{fig:relation_expansion}). He selects the relation "desires," leading to prompts like "extra service fee" and "identify nearby attractions." To further develop his idea, Alice uses the \textit{query-based idea expansion} feature (Figure \ref{fig:note_query}) to generate additional thoughts on "the downsides of living with Airbnb," and adds a new note about "infrequent room cleaning service" to the whiteboard using the \includegraphics[height=1em]{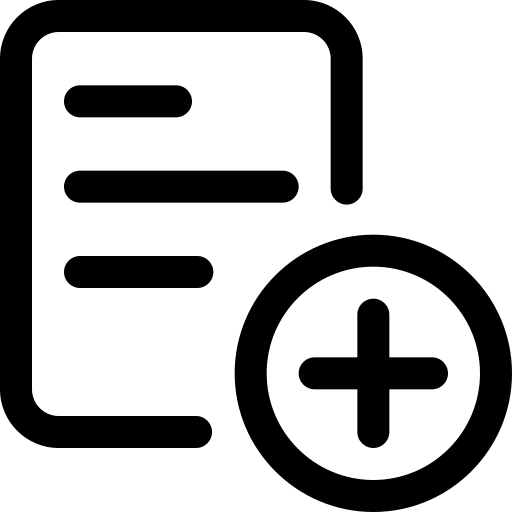} button.

As the team accumulates more ideas on the display, they decide to organize them to better discern themes and shared interests. They click the "Group" button on the workspace menu and choose "Global grouping" (Figure \ref{fig:grouping_process}). LADICA then offers multi-faceted affinity lenses for categorizing the ideas from various perspectives. The team opts for an affinity lens focused on "planning and preparation aspects of the ideas," leading LADICA to sort the ideas into groups like "Accommodation," "Local Transportation," and "Financial Consideration." Interested in delving deeper into "Accommodation," Alice and Bryan select \includegraphics[height=1em]{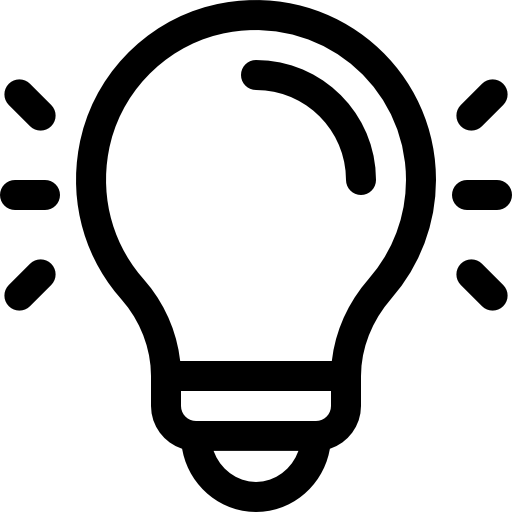} on the group sidebar to prompt LADICA for discussion hints. They also click \includegraphics[height=1em]{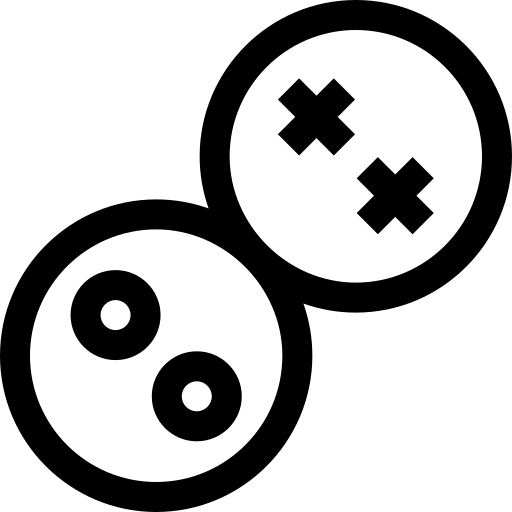} to perform a hierarchical grouping analysis within the "Accommodation" category.

During their group brainstorming session, the team activates the \textit{relation hint} feature (Figure \ref{fig:relation_hint}) to ensure everyone understands how their ideas connect. LADICA updates these relationships to improve mutual awareness. Simultaneously, they record their informal discussions (Figure \ref{fig:dialogue_feature}), allowing LADICA to automatically synchronize the workspace content with their conversation, identify key information relevant to their task, and retrieve related existing ideas.

\subsection{Key Features}
\label{sec:key_features}

In this section, we introduce the key features in each layer that were demonstrated in the example scenario.

\subsubsection{Idea repository layer}
\label{sec:idea_layer}

This layer aims to provide cognitive assistance for group ideation process and address the cognitive and social-motivational challenges identified in the formative study. The features in this layer includes group goal decomposition, query-based and relation-based idea expansion and key information retrieval from live discussion.


\begin{figure*}[t!]
  \centering
  \includegraphics[width=\linewidth]{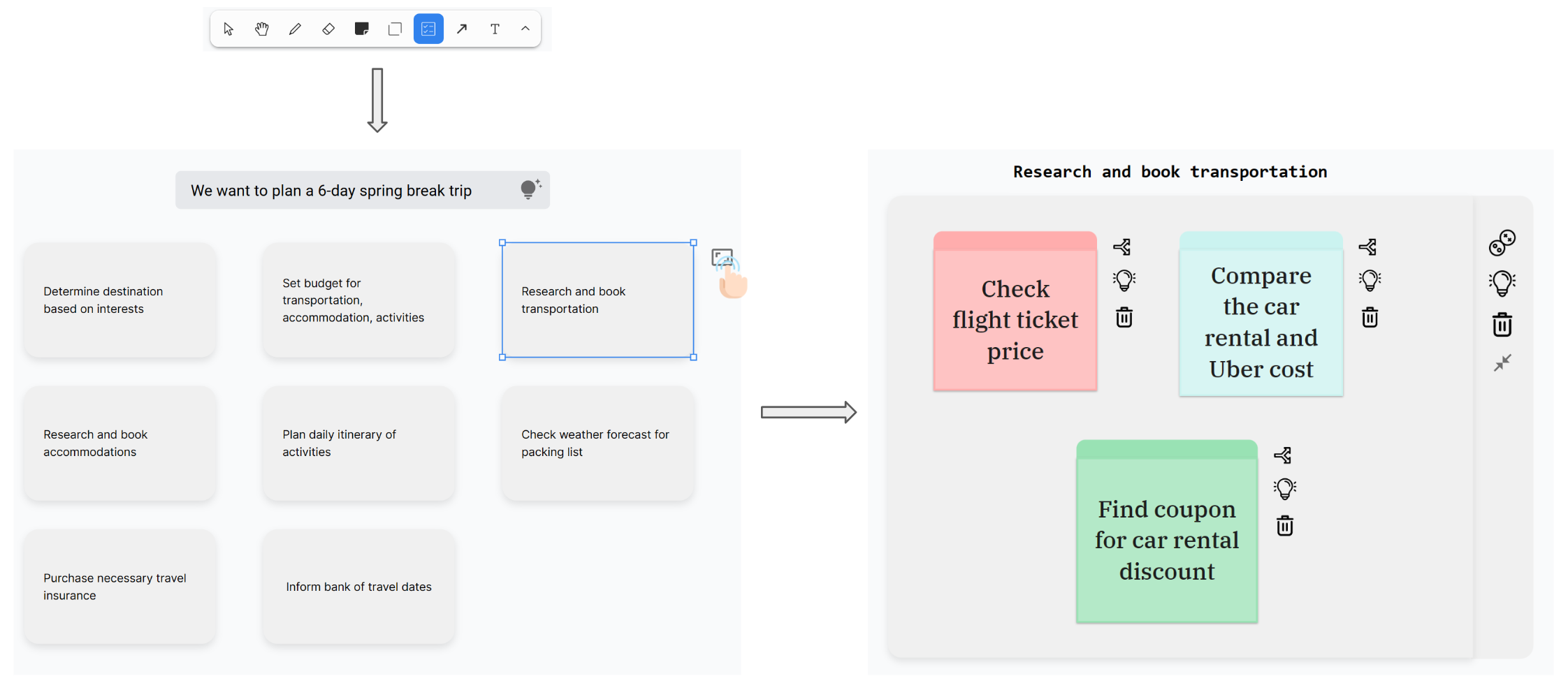}
  \caption{The interaction flow of group goal decomposition. The team can activate the feature by pressing \includegraphics[height=1em]{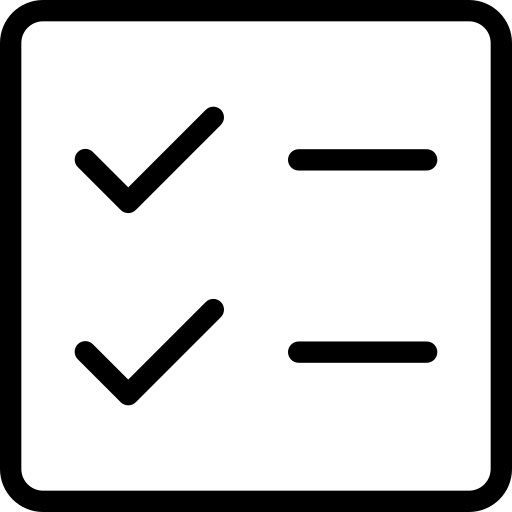} button on toolbar. Then they can enter their group goal on query bar and LADICA will suggest sub-goals for group collaboration. The team can then press \includegraphics[height=1em]{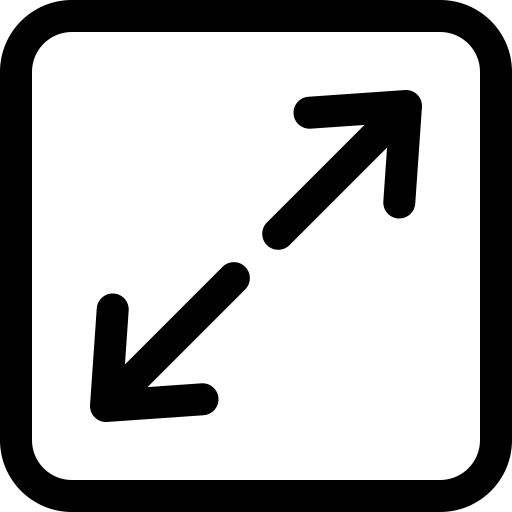} button to expand the sub-goal into a topic group for group ideation.}
 
  \label{fig:goal_decomposition}
  \vspace{-3mm}
\end{figure*}

\paragraph{\textbf{Facilitating groups to decompose and identify task structure}}

Understanding the structure of a group task is essential for creating an initial shared mental model that assists teams in coordinating their efforts on specific sub-goals (DG1), especially when teams conduct a top-down group ideation process. To facilitate this cognitive process, LADICA provides the \textit{group goal decomposition} feature that prompts the division of an overarching group goal into more manageable components (Figure \ref{fig:goal_decomposition}). This feature is initiated when groups enter their task goal into the task decomposition query bar. By pressing the ``generate subtasks'' button, LADICA generates cards that suggest the subtasks associated with the overall team goal. We instructed the LLM to provide just enough detail for the subtasks, avoiding excessive elaboration that could dominate the team's thought process on how to approach each subtask. The users can unfold a subtask card into a group subspace by pressing the \includegraphics[height=1em]{figure/maximize.png} button and exchange their ideas about the subtask within the workspace.


\paragraph{\textbf{Supporting articulation and expansion of team idea}}

\begin{figure*}[t!]
  \centering
  \includegraphics[width=\linewidth]{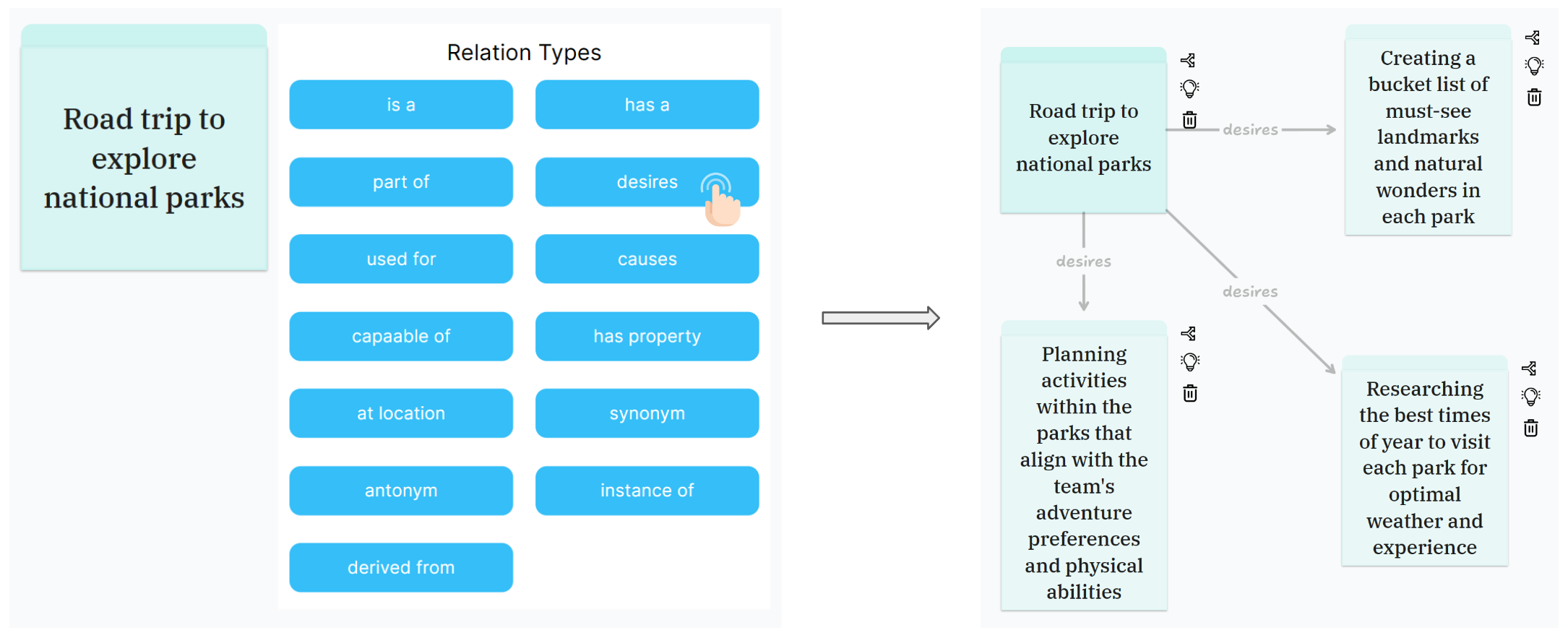}
  \caption{The interaction flow of relation-based expansion of an idea. Users can activate this feature by pressing the \includegraphics[height=1em]{figure/arrow.png} button next to an idea note. Then, they can choose a relation type to explore, and LADICA will generate potential thinking aspects associated with the idea based on the selected relation type.}
 
  \label{fig:relation_expansion}
  \vspace{-3mm}
\end{figure*}

\begin{figure*}[t!]
  \centering
  \includegraphics[width=\linewidth]{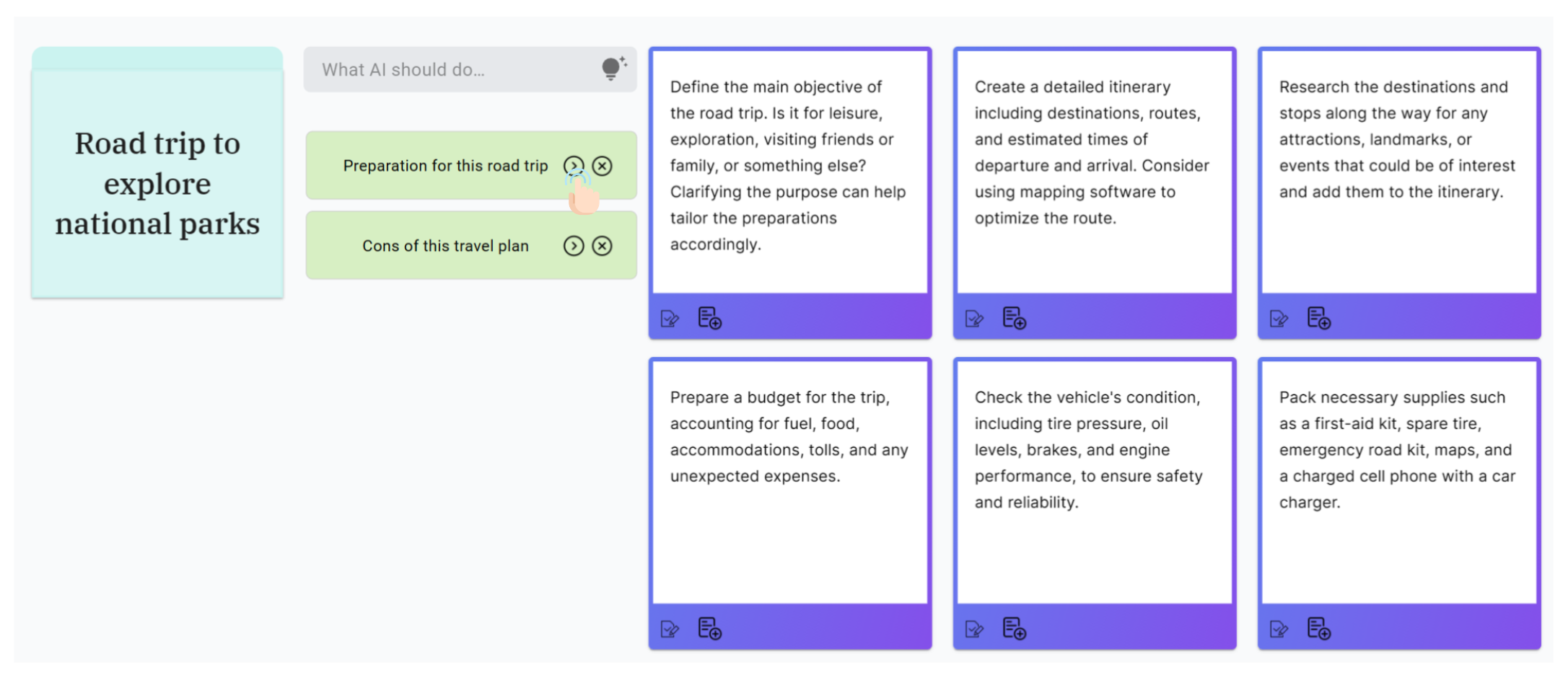}
  \caption{The interaction flow of query-based expansion of an idea. Users can activate this feature by pressing \includegraphics[height=1em]{figure/idea.png} button next to an idea note. They can then enter a query, and LADICA will provide a list of thinking direction hints based on the user's query. Users can press \includegraphics[height=1em]{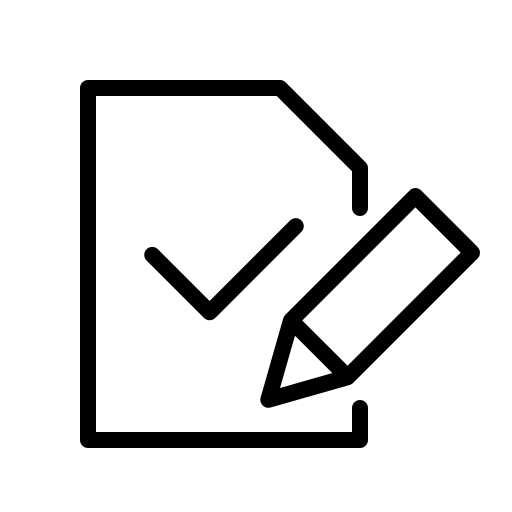} to automatically revise the idea based on the hint or \includegraphics[height=1em]{figure/add.png} to add it as a new note for further exploration. }
 
  \label{fig:note_query}
  \vspace{-3mm}
\end{figure*}

Our formative study revealed the user needs of articulating and expanding their ideas effectively in co-located team collaboration (DG1). LADICA offers two types of cognitive support to ignite a team's deep and diverse thinking about their initial ideas. The \textit{query-based expansion} about an idea (Figure \ref{fig:note_query}) enables users to discover potential directions to further develop their chosen idea. Users can press \includegraphics[height=1em]{figure/apply_suggestion.png} to automatically revise the idea based on the hint or press \includegraphics[height=1em]{figure/add.png} to add the selected hint as a note for further discussion and exploration. 

Furthermore, LADICA enables \textit{relation-based expansion} of an idea (Figure \ref{fig:relation_expansion}). By pressing \includegraphics[height=1em]{figure/arrow.png} along with an idea note, users can choose a specific type of relationship and ask the system to provide suggestions on topics they can explore following this relation. It is important to note that LADICA's primary focus is on hinting at thinking directions rather than directly generating ideas. By providing these cognitive supports, LADICA aims to facilitate teams' understanding of how to articulate and expand their ideas while still requiring individual or collective effort to produce specific ideas. This approach strikes a balance between system-assisted inspiration and user-driven ideation, fostering a collaborative environment that encourages deep thinking and diverse perspectives.



\begin{figure*}[t!]
  \centering
  \includegraphics[width=\linewidth]{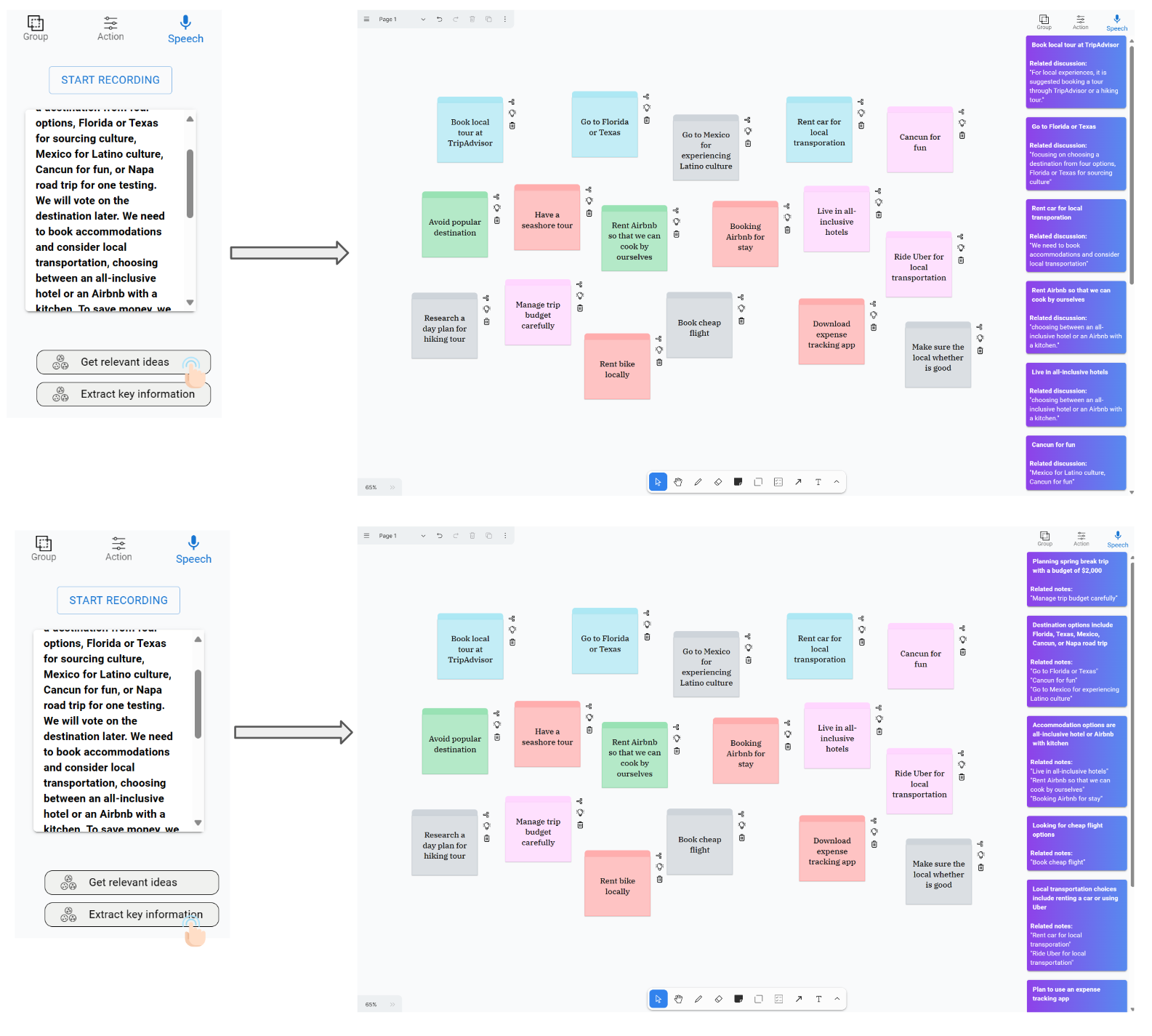}
  \caption{The interaction flow of discussion-based key information extraction and retrieval. Users can start recording the ongoing discussion by pressing ``Start Recording'' under the ``Speech'' menu. The live transcription is displayed simultaneously. By pressing ``Get relevant ideas,'' LADICA will identify existing ideas related to the discussion, \zzrevision{where each card on the right showing an existing idea relevant to the ongoing discussion and corresponding transcription}; By pressing ``Extract key information'', LADICA will extract key information from the discussion and also highlight related existing idea notes, \zzrevision{as shown in cards on the right in the lower figure.
  }}
 
  \label{fig:dialogue_feature}
  \vspace{-3mm}
\end{figure*}

\paragraph{\textbf{Extracting key information from ongoing discussion}}

During the ideation stage, co-located teams typically share their ideas verbally, which can add extra cognitive load when they need to remember and transfer the key information from their discussion onto an external shared workspace \zzrevision{(DG2)}. To alleviate this inconvenience, LADICA offers a \textit{discussion-based key information extraction} feature that allows teams to focus on discussing ideas while automatically highlighting key information once the discussion ends. As shown in Figure \ref{fig:dialogue_feature}, teams can press the ``Speech'' button on the workspace menu and select ``Start recording'' to record their discussion. LADICA will display the latest transcription of their conversation as it progresses. When the team concludes a section of their discussion, they can press ``Extract key information'' to identify the key points from their dialogue. If a key point relates to existing idea notes on screen, LADICA will also show the related note in each information card. Teams can then press on a key information card to create an idea note containing the same content.




\subsubsection{Affinity lens layer}
\label{sec:affinity_layer}

\begin{figure*}[t!]
  \centering
  \includegraphics[width=\linewidth]{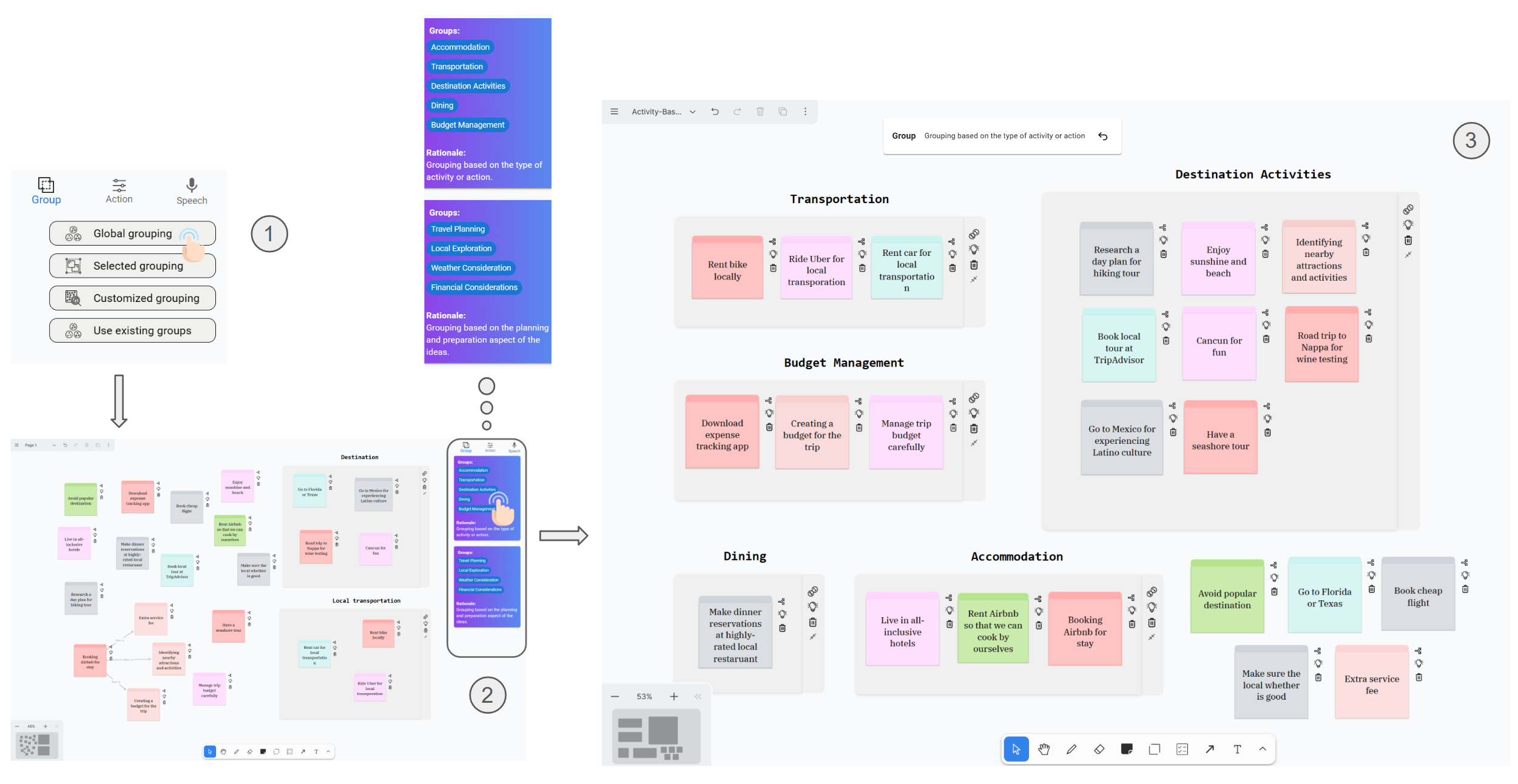}
  \caption{Illustration of the process of affinity-based grouping. In the first step, users can press the ``Group'' menu and choose a grouping approach. LADICA will then suggest multi-faceted affinity lenses for organizing the existing ideas. By selecting an affinity lens, LADICA will automatically group the ideas through the lens and display the results on a corresponding grouping view page. The rationale behind the grouping will also be shown on the new page. Users can navigate between view pages using the drop-down menu (Figure \ref{fig:teaser}, E).}
 
  \label{fig:grouping_process}
  \vspace{-3mm}
\end{figure*}

As described in Section \ref{sec:layers}, the affinity lens layer aims to facilitate the organization of shared ideas and support an exploratory grouping process that helps teams analyze their ideas from different perspectives (DG3). Based on the description of the KJ method~\cite{scupin1997kj} underlying affinity diagramming, we define the affinity lens as \textit{a specific viewpoint through which a varied collection of ideas can be organized into cohesive clusters, showcasing the commonalities or connections as perceived from that perspective}. In particular, this layer offers cognitive assistance including flexible creation of multi-faceted affinity lens, automatic grouping and tracking of different affinity lens, hierarchical grouping analysis and rollback to saved snapshot.

\paragraph{\textbf{Allowing flexible creation of multi-faceted affinity lens}}

Brainstorming different perspectives to analyze ideas requires team members to think critically and comprehensively. To facilitate this cognitive process, LADICA offers three complementary methods for hinting affinity lenses with different user needs. As depicted in Figure \ref{fig:grouping_process}, users can access affinity lens creation features by pressing ``Group'' button in the workspace menu. The \textit{global grouping }generates affinity lenses based on all ideas presented in the shared workspace. Additionally, users can include a subset of ideas for affinity analysis using \textit{selected grouping} or provide specific criteria for generating affinity lenses through \textit{customized grouping}. For each request, LADICA generates multiple affinity lenses that offer diverse analytical perspectives. Each affinity lens comprises a set of grouping affinities and a description of the affinity lens itself.


\paragraph{\textbf{Automating grouping and tracking of different affinity lenses}}

LADICA facilitates automatic grouping and tracking of affinity lenses to reduce the team's cognitive load associated with performing and updating categorization. As illustrated in Figure \ref{fig:grouping_process}, by selecting an affinity lens, LADICA will group ideas into suggested affinities and display the results on an additional workspace page. Users can switch between lens pages (Figure \ref{fig:teaser}, E) to review and compare groupings with different affinity lenses. All lens pages are synchronized with the main workspace page, meaning that edits made on the main workspace page will automatically apply to the others. When the team switches to a lens page, LADICA will automatically regroup the new ideas based on the corresponding affinity lens, facilitating tracking of the latest grouping results throughout the collaboration.


\begin{figure*}[t!]
  \centering
  \includegraphics[width=\linewidth]{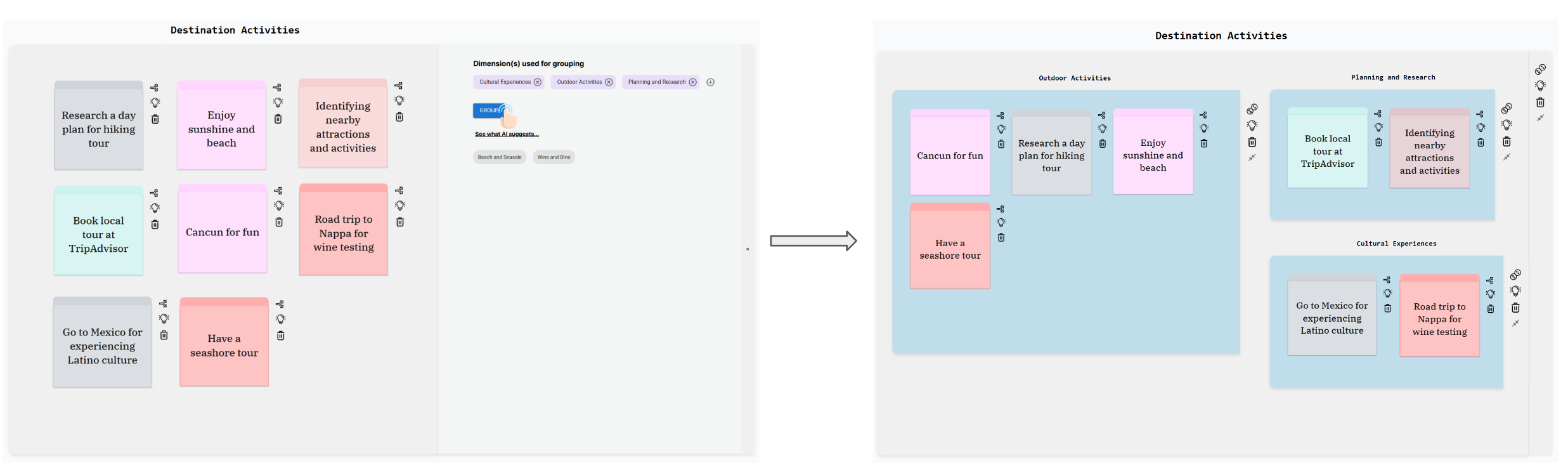}
  \caption{Illustration of the process of hierarchical affinity analysis. Teams can access the hierarchical grouping panel by pressing \includegraphics[height=1em]{figure/two_group.png} on the sidebar of a group. Users can either input the grouping dimensions themselves or request LADICA to suggest dimensions based on the group topic and children's ideas. By pressing the ``Group'' button, LADICA will organize the ideas within the group into different affinity sub-groups.}
 
  \label{fig:hierarchical_grouping}
  \vspace{-3mm}
\end{figure*}

\paragraph{\textbf{Allowing hierarchical affinity analysis}}

The meta-cognitive prompting theory~\cite{wiltshire2014training} suggests that providing teams with hierarchical prompts can facilitate their collective in-depth thinking and foster awareness of the hierarchical structure, aligning their work at more detailed levels. LADICA offers a \textit{hierarchical affinity analysis} feature (Figure \ref{fig:hierarchical_grouping}) that enables teams to perform flexible in-depth grouping within a particular group. To use this function, users can press  \includegraphics[height=1em]{figure/two_group.png}  (Figure \ref{fig:teaser}, D) on the group sidebar to view the hierarchical grouping panel for this group. Users have the option to input their own affinities or ask AI to generate potential grouping affinities for them by pressing the ``see what AI suggests'' button. Following this, by pressing the ``Group'' button, LADICA will categorize the ideas within the group based on the applied affinities. Users can move a subgroup outside its parent to make it a top-level group when they want to conduct further analysis.


\paragraph{\textbf{Allowing rollback to previous idea snapshot}}

When engaging in exploratory analysis of their ideas, co-located teams seek the flexibility to navigate through different snapshots of the workspace, allowing them to experiment with various groupings (DG3). LADICA offers a snapshot saving and loading function that enables teams to revert to previous key moments of the collaboration displayed on the screen. By pressing ``Action'' then ``save snapshot''  in the workspace menu, users can save the current workspace snapshot with a specified name. The snapshot will be stored in a Firebase database for future access. Users can press ``load snapshots'' and choose a previously saved snapshot to load.

\subsubsection{Discussion reference layer}
\label{sec:discussion_layer}

The discussion reference layer is designed to promote and support group discussions for generating new insights based on the ideas displayed on the screen. Specifically, this layer offers features such as relation hints among ideas, discussion-based idea retrieval, and group-based ideation hints.

\begin{figure*}[t!]
  \centering
  \includegraphics[width=\linewidth]{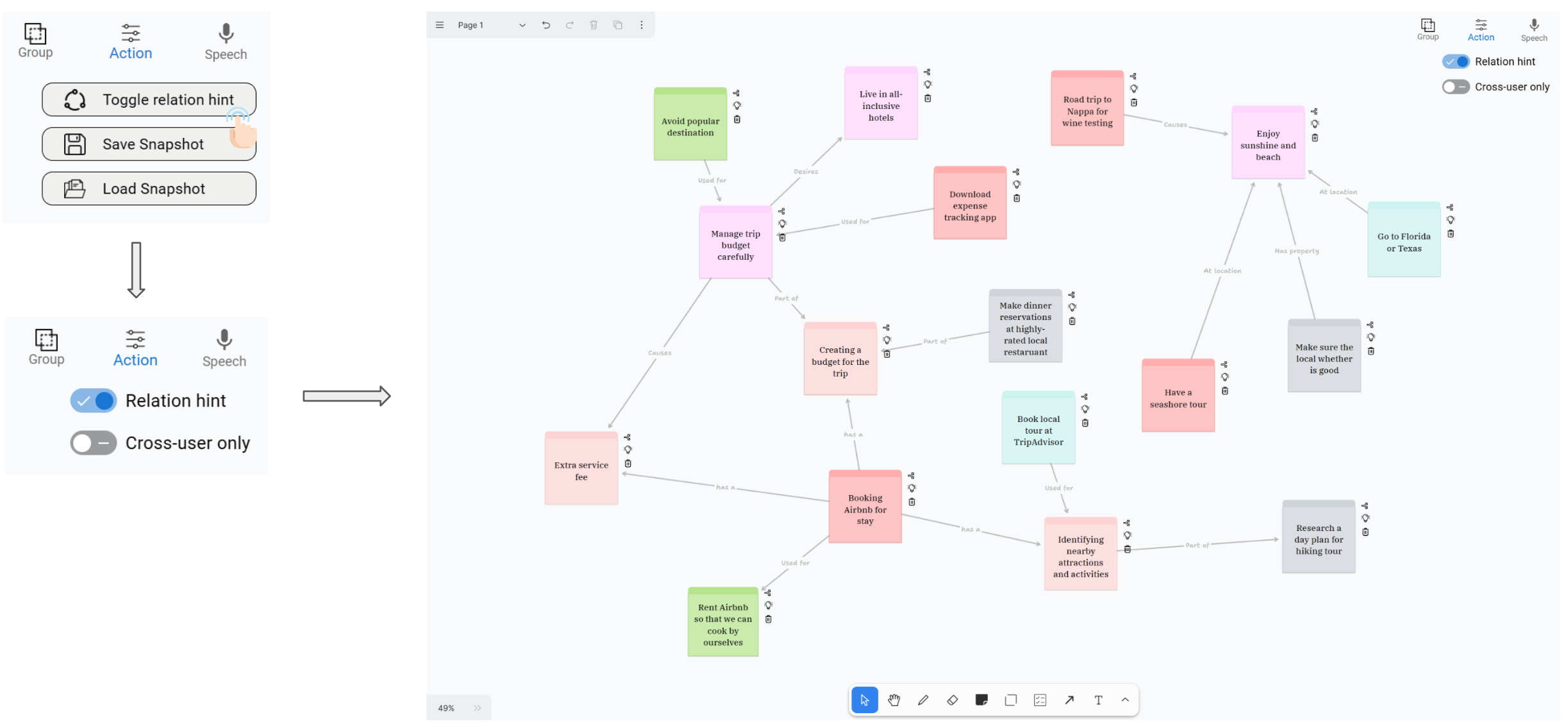}
  \caption{Illustration of the process of relation hint generation. Teams can toggle the relation hint feature under ``Action'' menu. Besides, they can also choose to only generate relations among ideas created by different teammates. Once activated, LADICA will periodically refresh the relation hints every ten seconds.}
 
  \label{fig:relation_hint}
  \vspace{-3mm}
\end{figure*}

\paragraph{\textbf{Enhancing mutual awareness for team members}}

Establishing and preserving mutual awareness is essential for team coordination and recognizing opportunities for collaboration. To support this meta-cognitive process, LADICA is able to provide relation hints among ideas and update them periodically throughout the collaboration \zzrevision{(DG4)}. As shown in Figure \ref{fig:relation_hint}, users can press ``Action'' then ``Toggle relation hints'' in the workspace menu to activate the relation hint generation. In addition, users can toggle the ``cross-user only'' to only generate ideas created by different team members. To reflect the relations among the latest ideas, LADICA will update the relation hints every 10 seconds when the function is activated. Note that this feature is designed to indicate relationships to catalyze initial discussion rather than to suggest specific discussion content. To achieve this, we instruct the LLM to generate relation types based on the ConceptNet relation taxonomy~\cite{speer2017conceptnet}, which encompasses a wide range of types and maintains a level of granularity that provides teams with ample room to explore  the relationships.

\paragraph{\textbf{Retrieving existing ideas relevant to ongoing discussion}}

Our findings indicate that team members often struggle to keep the whiteboard content aligned with the dynamic flow of their discussions. To address this challenge, LADICA is designed to intelligently retrieve relevant existing idea nodes based on the content of the conversation \zzrevision{(DG2)}. Users can activate this feature by pressing the ``Start Recording'' button at the beginning of their collaborative session (Figure \ref{fig:dialogue_feature}). This prompts LADICA to start recording while simultaneously analyzing the ongoing discussion to identify mentions or references related to existing ideas. Upon ending the session with the 'Stop Recording' button, LADICA processes the recorded conversation. The system then retrieves the existing ideas that have relevance score with its corresponding dialogue reference in the discussion over a predefined threshold (see Section \ref{sec:speech}). For each relevant ideas, LADICA also shows the corresponding dialogue reference.

\begin{figure*}[t!]
  \centering
  \includegraphics[width=\linewidth]{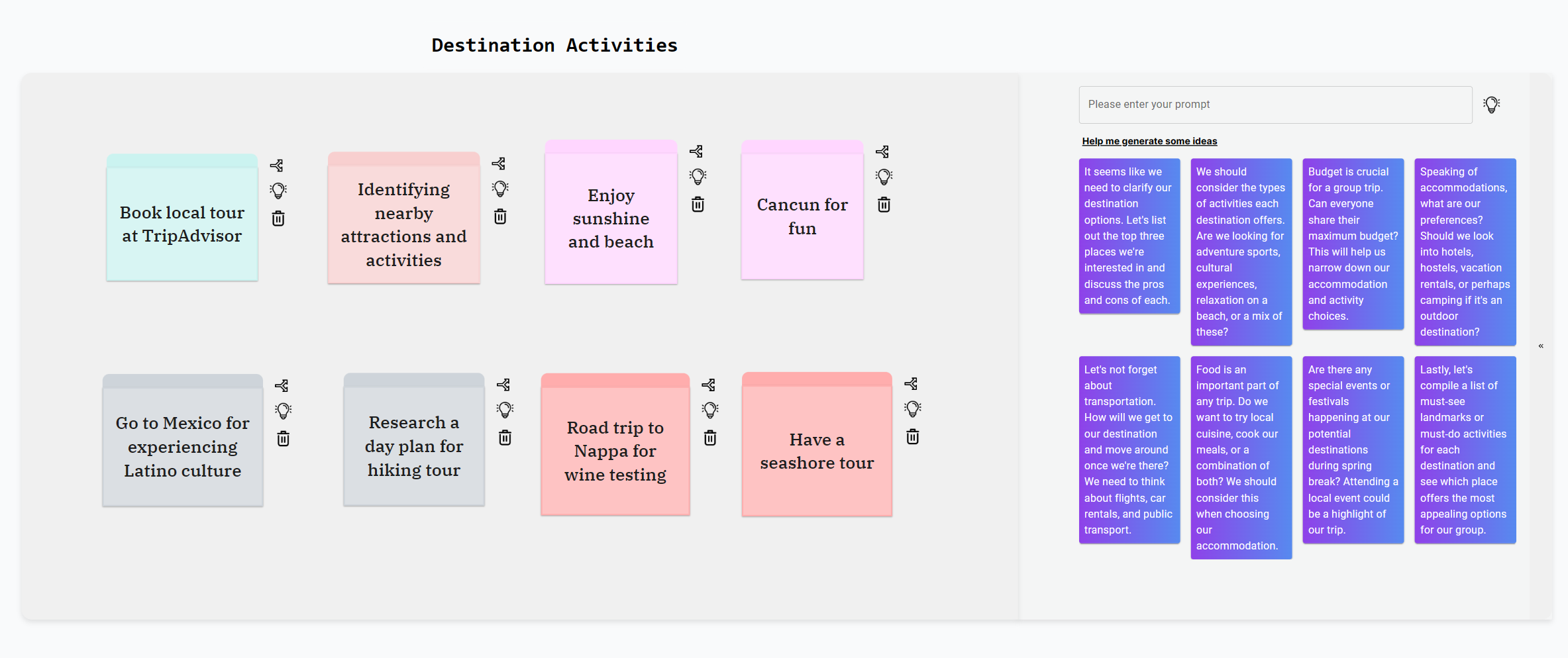}
  \caption{Illustration of the group-based discussion hints. Teams can access the feature by pressing \includegraphics[height=1em]{figure/idea.png} on the sidebar of a topic group. They can enter specific instructions for hint generation or press ``Help me generate some ideas'' to ask LADICA to automatically suggest some discussion hints based on the group content.}
 
  \label{fig:discussion_hint}
  \vspace{-3mm}
\end{figure*}

\paragraph{\textbf{Providing group-based discussion hints}}

A crucial opportunity for generating new insights is to reflect on team's ideas within group, which sparks diverse perspective in discussion. To facilitate this process, LADICA provides \textit{group-based discussion hint} (Figure \ref{fig:discussion_hint}) that is able to generate hints of discussion topics about the theme of the group and existing ideas within the group. Teams can access this feature by pressing \includegraphics[height=1em]{figure/idea.png} in the group sidebar (Figure \ref{fig:teaser}, D). Users can provide further instructions for hint generation. It is important to note that the hints are not intended to generate ideas directly; instead, they are designed to inspire potential new directions for team thinking and discussion \zzrevision{(DG4)}.


%


%

\subsection{Implementation}
\label{sec:implementation}

LADICA is a multi-user web application with a React-based\footnote{\href{https://react.dev/}{https://react.dev/}} front-end client, Flask-based\footnote{\href{https://flask.palletsprojects.com/}{https://flask.palletsprojects.com/}} back-end server and Firebase\footnote{\href{https://firebase.google.com/}{https://firebase.google.com/}} database. The AI features are implemented based on OpenAI's GPT-4 turbo API\footnote{\href{https://platform.openai.com/docs/models/gpt-4-and-gpt-4-turbo}{https://platform.openai.com/docs/models/gpt-4-and-gpt-4-turbo}}. The technical details are discussed in the following sections. The prompt for each feature is shown in Appendix.

\subsubsection{Collaborative whiteboard}
\label{sec:technical_whiteboard}

We utilize the \textit{tldraw}\footnote{\href{https://tldraw.dev/}{https://tldraw.dev/}} framework for creating the whiteboard application. Within this framework, we have implemented custom shape classes to visualize objects such as notes, groups, relations, search bars, and more. Note that most off-the-shelf browsers do not support multi-touch interactions on web applications, except for gesture purposes, even if the display is multi-touch enabled. To overcome this limitation and, we use \textit{yjs}\footnote{\href{https://yjs.dev/}{https://yjs.dev/}} library to allow real-time collaborative editing on a shared whiteboard workspace via personal touchable devices such as phones and tablets in addition to using the large display. This approach offers three additional advantages: (1) It enables the system to track different user identities, allowing for the implementation of personalized features such as individual coloring and cross-user collaborative hints; (2) It allows each user to freely move, zoom, and edit on the shared whiteboard canvas without disrupting the work of others; (3) This configuration enables easy deployment and supports most large touch screens equipped with a built-in browser or the function to mirror an external device's screen.

\subsubsection{Relational hint generation}

We employ the GPT-4 turbo model to generate relational hints between notes, adopting the relation types from ConceptNet~\cite{speer2017conceptnet} as our candidate types. The model is tasked with predicting the existence of a relation between each pair of notes and identifying the most likely relation type if one existed. During the pilot study, we observed that the ``\textit{related to}'' type in ConceptNet often led the model to predict overly general relations among the notes. To address this, we exclude it from our candidate types and instructed the model to provide an explanation for each predicted relation, along with a confidence score ranging from 0 to 1. Based on empirical experiments, we filter out relations with a confidence score below 0.6.

\subsubsection{Affinity diagramming generation}

As described in Section \ref{sec:affinity_layer}, our affinity diagramming generation consists of two steps. In the first step, we use GPT-4 turbo model to recommend a list of affinity lenses given selected notes and optional user instruction. For each affinity lens, we ask the model to return a brief description of the lens and the groups contained in the lens. To avoid the model proposing groups that are semantically similar or overlapping, we guide the model to self-assess the semantic similarity score between each pair of groups. The model iteratively refines the names of the groups until all similarity scores fall below 0.6.

When a user selects a specific lens, we prompt the model to categorize the selected notes into existing groups based on their thematic similarity to each group description. In line with traditional affinity diagramming practices, each note is assigned exclusively to one group. The regrouping function is activated whenever new notes are added and users opt to apply an existing lens to the updated notes.

\subsubsection{Ideation and discussion hint generation}
\label{sec:idea_generation}

We use the GPT-4 turbo model to generate ideation and discussion hints, as described in Sections \ref{sec:idea_layer} and \ref{sec:discussion_layer}. For \textit{query-based idea expansion}, we instruct the model to act as ``\textit{an intelligent group facilitator that can expand the team's thinking on their ideas by offering thinking hints based on user requirements}''. We request the model to self-evaluate each candidate hint and assign a confidence score from 0 to 1, only returning those with a score above 0.6. Similarly, for group-based discussion hint generation, we ask the model to function as ``\textit{a group facilitator that provides further hints to stimulate group discussion based on user instructions, group topic, and ideas within the group}''. We also guide the model to particularly consider the differences and commonalities among different users' ideas when generating discussion hints.

\subsubsection{Speech recognition with information retrieval and extraction}

\label{sec:speech}

LADICA uses OpenAI's Whisper model\footnote{\href{https://platform.openai.com/docs/guides/speech-to-text}{https://platform.openai.com/docs/guides/speech-to-text}} to transcribe conversation recording into text. With audio transcription, we implemented the \textit{key information extraction} feature through instructing GPT-4 model to find key information in the conversation. Drawing inspiration from Pask's conversation theory~\cite{pask1976conversation}, our focus is on finding key arguments, evidence, ideas, actions, activities, and concepts in the dialogue. Additionally, the model is asked to assign a relevance score between 0 and 1 to each potential key point in relation to the existing notes on the whiteboard. Candidates with a score below 0.6 are discarded. To identify existing notes that are relevant to the ongoing conversation, we instruct the GPT-4 model to first determine the most relevant sentence in the transcription for each note, along with a relevance score ranging from 0 to 1. We then retrieve notes that have a relevance score greater than 0.6.
\section{User Study}

To evaluate the usability, effectiveness, and usefulness of LADICA, we conducted a lab user study with 14 participants. The study aims to answer the following research questions:

\begin{itemize}
    \item \textbf{RQ1}: Can LADICA successfully integrate into and support ideation and planning tasks for collocated groups?
    \item \textbf{RQ2}: How effective is LADICA in supporting group ideation, organization, and discussion during team collaboration?
    \item \textbf{RQ3}: How does the AI assistance introduce by LADICA affect the collocated group collaboration?
    \item \textbf{RQ4}: What challenges do users face while using LADICA in collaborative group tasks?
\end{itemize}

\subsection{Participants}

We recruited 14 participants (5 males, 9 females; average age=23.9) from a university mailing list. Each participant received a \$15 USD gift card as compensation. The group consisted of nine undergraduate and five graduate students with diverse academic backgrounds, including computer science, business, anthropology, and social sciences. All participants had prior experience with collaborative work in teams and were familiar with using both traditional and digital whiteboard tools, such as Miro and Lucidspark, for collocated group tasks. Furthermore, they all had experience with Large Language Models (LLMs) such as ChatGPT, Gemini, and Claude.

\subsection{Apparatus}
\label{sec:apparatus}

\begin{figure}[t!]
  \centering
  \includegraphics[width=\columnwidth]{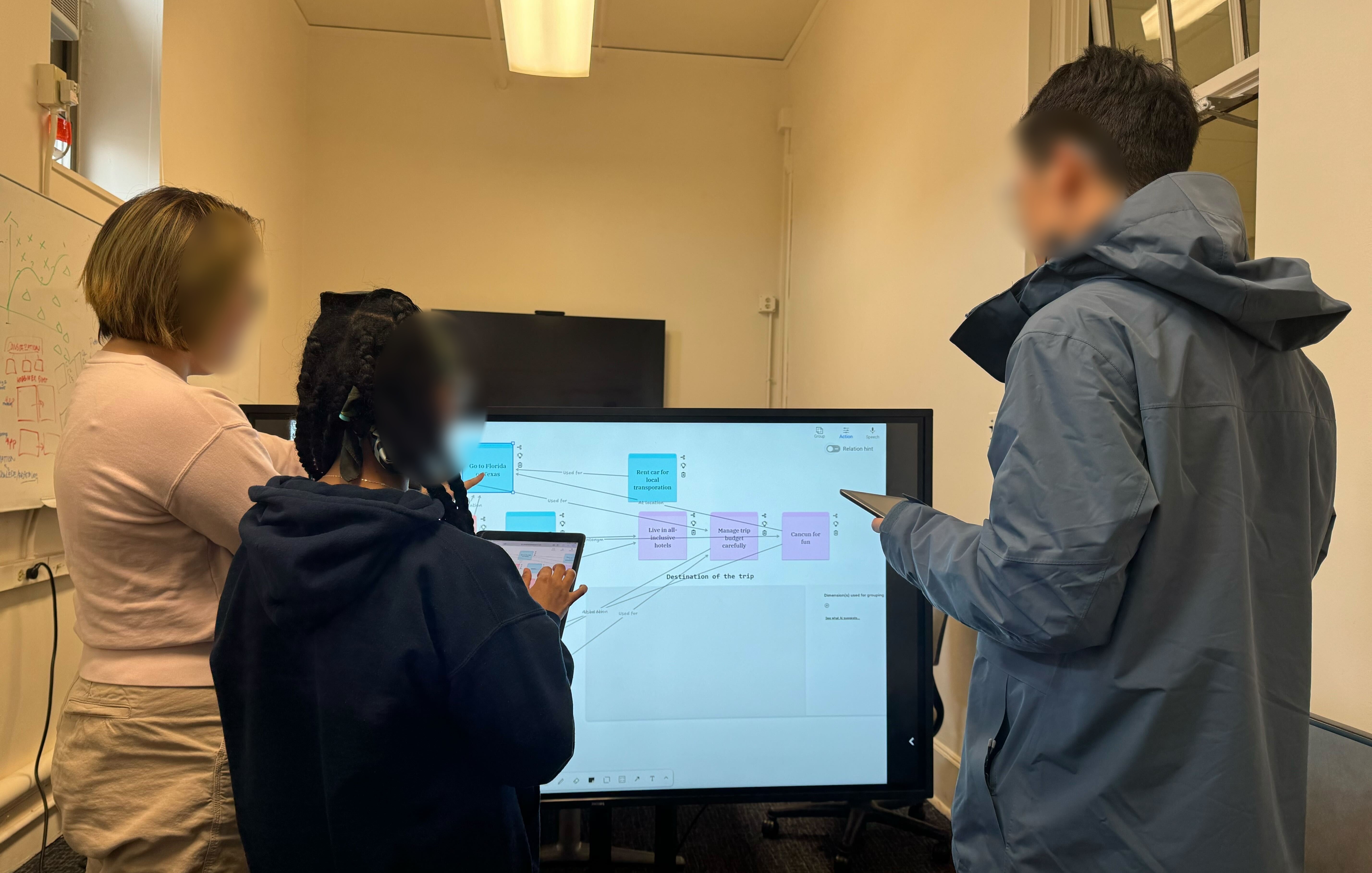}
  \caption{The illustration of user study setting}
 
  \label{fig:study_setting}
  \vspace{-3mm}
\end{figure}

The study took place in-person within a usability lab. The participants were organized into small groups, each consisting of three or four individuals. LADICA ran on a 65-inch Phillips multi-touch display with an integrated Chromium browser. As discussed in section~\ref{sec:technical_whiteboard},each participant was also
provided with an iPad to enable simultaneous content creation (especially text entry) and editing. All iPads and the shared touch display were logged into the same workspace in LADICA, ensuring content synchronization. This setup allowed for all actions performed on the large screen or on any individual iPad to be visible to all members of the group. \zzrevision{We also set up an external video camera to record the study sessions}. Figure \ref{fig:study_setting} illustrates the study apparatus setting. 


\begin{figure*}[t!]
  \centering
  \includegraphics[width=\linewidth]{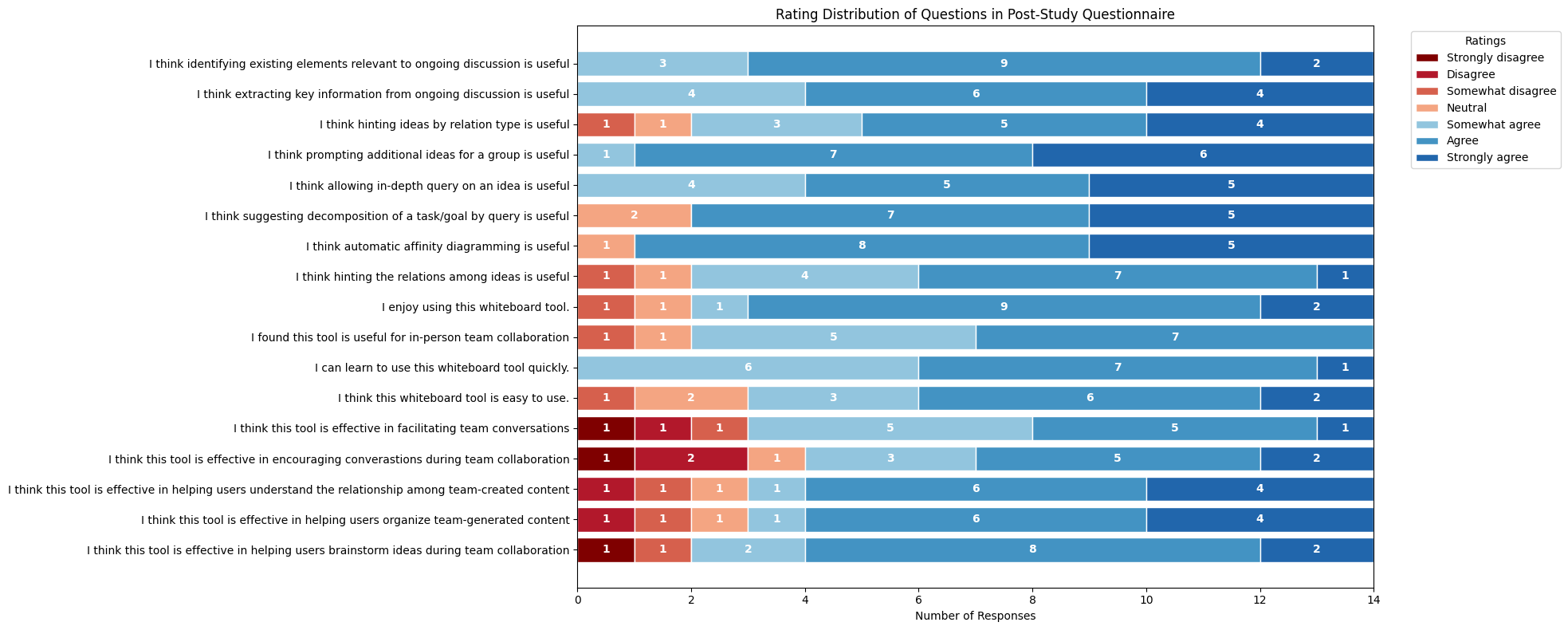}
  \caption{Results from the post-study questionnaire about usability, usefulness and effectiveness of LADICA}
 
  \label{fig:study_result}
  \vspace{-3mm}
\end{figure*}

\subsection{Study Procedure}

Each study session involved a group of three or four participants working together in person to complete designated tasks, using the setup described in Section \ref{sec:apparatus}.

\subsubsection{Tasks}
The groups were assigned two tasks in random order, designed under two guiding principles: (1) the tasks do not require specialized knowledge, and (2) they reflect common types of cognitive activities found collocated team collaboration tasks.\looseness=-1

\paragraph{Group trip planning}

The task of planning a group trip required participants to identify key components (like destination, accommodation, and transportation) and brainstorm ideas for each.  When working on different parts simultaneously, teams must consider how their section relates to others. It requires team members to communicate with each other to align their interests and develop a plan that meets everyone's needs. For example, members planning the destination or transportation options for a specific day must coordinate with those arranging accommodation for the same day. 

The task instruction we gave the participants was as follows: ``\textit{Imagine you are planning a Spring break trip together. Your objective is to brainstorm, negotiate, and devise a detailed plan covering key aspects such as travel, accommodation, activities, and budgeting using the system. Please adhere to the following constraints for the trip: (1) Each participant has a budget limit of approximately \$2,000; (2) The trip lasts for 7 days}''.

\paragraph{School policy discussion}

The task focused on discussing a school policy demanded that teams conduct a thorough critical analysis, weighing both the advantages and disadvantages, and forecasting the policy's potential implications. Teams were expected to utilize multiple analytical perspectives and come up with comprehensive dimensions for the task. This issue typifies a ``wicked problem'' characterized by significant social complexity, the absence of clear metrics for success, and the understanding that solutions are not strictly right or wrong but rather vary in effectiveness based on multiple factors.

The instruction of this task was as follows: ``\textit{The flipped classroom model is an innovative educational approach that reverses the traditional learning environment by delivering instructional content outside of the classroom.... Imagine a school is contemplating the widespread adoption of the flipped classroom model across its campus. The school wants to know students’ feedback about this policy. Please use the system to support your discussion of the advantages and disadvantages of this policy, and how this policy would impact the student's overall school experience}''. 

\subsubsection{Procedure}

Each study session lasted 65 minutes, and each session consisted of four steps:

\textbf{Introduction (5 minutes)}
The researchers introduced the goal of the study. Participants introduced themselves, provided informed consent, and filled out a pre-study questionnaire.

\textbf{System Training (10 minutes)}
The researchers provided a tutorial of LADICA's features. After that, participants had the opportunity to familiarize themselves with these features, with the research team on standby to clarify any queries about the system's operation.

\textbf{Team Tasks (35 minutes)}
The main activity involved participants engaging in the previously described tasks. Each task was allocated 15 minutes with a 5-minute interlude for a break.

\textbf{Survey and Interview (15 minutes)}
The participants completed a post-study questionnaire designed to evaluate the system's perceived usefulness, effectiveness, and usability of the system to facilitate collocated group work. In a follow-up semi-structured interview, we aimed to gather participants' insights on how LADICA impacts and integrates into their collocated team collaboration, as well as their thoughts on its limitations and suggestions for system enhancement.

\subsection{Results}

\subsubsection{Post-study questionnaire}

Figure \ref{fig:study_result} presents a summary of the responses from the post-study questionnaire completed by 14 participants. Generally, the feedback confirmed the usability of LADICA, highlighting that its design and system features effectively facilitate group ideation, idea analysis, and discussion during co-located team collaboration. For any ratings that were below Neutral, participants were asked to elaborate on their reasons in the subsequent interviews. These detailed insights are discussed in Section \ref{sec:interview_result}.

\subsubsection{\zzrevision{Interaction Log Analysis}}

\zzrevision{We applied the interaction flow analysis method~\cite{avouris2003analysis} to analyze the interaction log data, conversation transcripts, and video recordings from study sessions, The goal was to understand how participants used LADICA during collaboration and to assess the fulfillment of the design goals. One of the authors reviewed the qualitative data to identify and categorize interaction patterns within the system.}


\paragraph{\zzrevision{Ideation and knowledge sharing}}

\zzrevision{LADICA offers features such as \textit{group goal decomposition, query-based idea expansion, and relation-based idea expansion} to encourage dynamic ideation and knowledge sharing (DG1). We observed that the participants used the group goal decomposition seven times throughout all study sessions. four of these instances occurred during the early discussion phase, where team members aimed to understand relevant topics or break down subtasks for the initially assigned topic (e.g., P4: ``\textit{Let's see what we need to consider for planning a spring break trip to Miami}''). The remaining instances took place in the middle phase of collaboration, when participants focused on diving into topics of shared interest, such as ``\textit{engaging students for in-class discussion}'' or ``\textit{group budget management strategy}''. After viewing the results of goal decomposition, participants tended to review the recommended subtasks sequentially, either exploring a specific topic collaboratively or delegating tasks to individual members.}

\zzrevision{Participants used query based idea expansion 18 times and relation based idea expansion 15 times throughout the sessions. When an individual triggered the idea expansion features, in 19 of the 33 instances, they sought input from other group members to decide which subtask to explore together. In the remaining cases, participants worked independently on creating notes under the subtasks and later brought them to the group discussion. We also observed that in three instances, participants dismissed the suggested subtasks and chose to list the subtasks themselves, as the generated suggestions were either too generic or included trivial details. This highlights the need for further fine-tuning the prompt strategies to guide the model toward an appropriate level of granularity.}

\paragraph{\zzrevision{Streamline discussion and workspace synchronization}}

\zzrevision{LADICA offers discussion-based key information extraction and discussion-relevant idea retrieval to facilitate synchronization between discussion and workspace (DG2). We found that these two features were frequently used during the study sessions. All groups in the study activated the discussion-based key information extraction feature at the beginning of their task sessions. They used on this feature to automatically create notes for tracking key information, allowing them to focus on discussing the topic (e.g., P8: ``\textit{we can use the system to help us track key ideas in our conversation}''). However, participants observed that the system sometimes extracted insignificant information from their conversations, which they had to manually delete. This indicates the need for further refinement of the corresponding prompts and the addition of a filtering function. }

\zzrevision{All of the groups used \textit{discussion-relevant idea retrieval} to keep track of relevant existing ideas while engaging in discussion. They examined the system-prompted ideas after having a conversation on a specific topic and identified connections between the existing ideas and the newly discussed points. Group 2 also alternated between using key information retrieval and extraction to efficiently synchronize discussion and workspace during their collaboration. This demonstrates the usefulness of these features in helping groups connect ongoing discussions with existing content.}

\paragraph{\zzrevision{Diverse analytical perspectives}}

\zzrevision{The affinity lens layers offer various functions to enable diverse and comprehensive perspectives for analyzing created content (DG3). Participants used the multi-faceted affinity lens generation feature a total of seven times during the study sessions. They employed this feature to either identify common themes among their divergent ideas generated during individual brainstorming or to discover additional perspectives to explore based on existing ideas. After the affinity lens was generated, groups tended to focus on a small subset of lenses that the entire group was interested in and extended their discussion by enabling hierarchical affinity analysis on the sub-topics identified. We found that two groups used the idea snapshot rollback feature to return to a previous state, as they found that the discussion had become too narrow or had diverged from the main topics they were assigned. We also observed that Group 1 did not use the affinity lens feature after trying it once because members felt that identifying common themes on their own would provide an opportunity for collaboration and help develop mutual understanding.' }

\paragraph{\zzrevision{Collaborative cues}}

\zzrevision{LADICA fosters group discussion by offering relation hints among ideas and group-based ideation prompts (DG4). During our study, participants used relation hints 10 times to gain insight into the relationships between different members' contributions. We observed that highlighting these relationships motivated group members to discuss the connected ideas and collaborate to expand on them together. However, since the system could present an overwhelming number of relations among ideas created by multiple members, some participants chose not to use this feature again, as they felt it might disrupt others' ongoing work and clutter the interface. This suggests the need to refine the number of hints provided at a time. Participants activated group-based ideation hints 12 times throughout the study sessions, primarily using the feature when they identified topics of common interest and sought prompts for further discussion. In five instances, participants directly included the suggested ideas, while in the remaining seven cases, they used the hints as inspiration to create their own ideas. }

\subsubsection{Post-study interview}
\label{sec:interview_result}

Using established reflexive thematic analysis method~\cite{braun2012thematic}, we coded interview transcripts and identify the following key findings into how users interact with and experience LADICA.

\paragraph{KF1: Dynamic Roles of LADICA in Co-located Team Collaboration}

Participants emphasized that LADICA served various roles during team collaborations. During individual brainstorming phases, the system acted as a \textbf{source of inspirational stimuli}, aiding users in rapidly expanding their ideas for further discussion within the team.  For example, P4 said ``\textit{I used the query and relation-based idea exploration to expand on my initial ideas and quickly come up with more to discuss with my team}''. Besides, P7 saw the ideation scaffolding features as a ``\textit{\textbf{bootstrap for team collaboration}}'', as they ``\textit{helped reduce solitary time, leaving more time for team discussions}''. Conversely, in the collective analysis phase, which requires deeper reflection, LADICA was seen as a \textbf{facilitator} for identifying differences or similarities in ideas. For example, P9 said ``\textit{we had diverse views on the flipped classroom model, but the system helped us identify the perspectives from which we could compare our ideas and extend our discussion in a meaningful way}''. Additionally, during active team discussions, LADICA's features acted as a \textbf{reminder} for the team, ensuring continuity and relevance in conversations. For example, P5 said ``\textit{the information extraction feature reminded us of key points in our ongoing discussions, and identifying relevant existing ideas helped us connect our current conversation to past ones}''.

\paragraph{KF2: Impact of Ideation and Relational Hints on Thinking and Conversation}

The participants appreciated the ideation and relational hints for their role in augmenting thought processes and conversations. The \textbf{query-based idea expansion} feature allowed them to ``\textit{easily know how to refine and extend an idea from an angle I (they) care about}'' (P10). Plus, the relation-based idea expansion feature helped user's \textbf{divergent thinking process}, as P12 said, ``\textit{It enables me to begin with a rough idea and delve deeper step by step. Moreover, it allows me to effortlessly explore a broad spectrum of related thinking directions}''. Features such as topic-based ideation hints and relational hints between ideas were also seen as instrumental in encouraging \textbf{inclusive participation} by all team members in discussions.  For example, P7 said ``\textit{the feature that offered additional ideation directions based on the topic we were discussing and our existing ideas aided us in uncovering new topics for discussion}''. P11 believed that ``\textit{understanding the relationships between our (their) ideas motivated me (him) to engage in discussions with others based on the dependencies among our (their) interests}''.

Nonetheless, there were concerns about the system potentially generating \textbf{too many relational hints} as the number of ideas on the whiteboard grew. This would ``\textit{make the whiteboard messy and increase our (their) cognitive load of understanding those relationships}'' (P6). Similarly, they also found sometimes ``\textit{an abundance of topic-based ideation hints makes it challenging to select the most interesting one for discussion}'' (P2). 

\paragraph{KF3: Grouping and Task Decomposition Facilitate the Construction of Shared Mental Models}

Participants found the features for task decomposition and grouping instrumental in fostering \textbf{a shared mental model} among team members during collaboration. At the beginning of the collaboration, the team goal decomposition feature allows team to ``\textit{immediately figure out what aspects and elements we (they) can think and discuss}'' (P5) and ``\textit{acts as an icebreaker for a team of strangers}'' (P1). P6 thought ``\textit{it provided us (them) with a common framework for thinking, allowing us (them) to contribute in areas of interest. By observing the group's focus, we (they) can easily understand the topics our teammates are considering}''.

Furthermore, the system's affinity-based grouping allowed for the easy \textbf{identification of commonalities among ideas} and facilitated their \textbf{comparison}. The ability to group ideas through multiple diverse lenses helps them ``\textit{understand the similarities and differences of their ideas from different perspectives}'' (P12). This promotes discussions that ``\textit{help build common ground}'' (P6). Leveraging the similarities suggested by the affinity groups, participants can ``\textit{more easily merge and connect relevant ideas compared to performing pairwise comparisons on an unstructured whiteboard}'' (P13).

\paragraph{KF4: Information Extraction and Retrieval Align Team's Internal and External memory in Live Discussions}

Participants acknowledged the significant role of discussion-based information extraction and retrieval features in \textbf{bridging their internal memory with the external shared workspace}.  P4 thought ``\textit{when we (they) are in the midst of a heated discussion, it's easy to forget what we (they) just said and lose track of some crucial points that have been raised. The information extraction feature helped me (her) recover those memories}'. Plus, P7 said ``\textit{it (information extraction) allows us (them) to focus on the discussion itself and frees us from the distracting task of transferring the content onto the whiteboard}''.

Moreover, the ability to retrieve ideas relevant to the ongoing discussion was seen as essential for \textbf{integrating past contributions with current deliberations}. For instance, P3 noted, "\textit{It helped me quickly connect what we've previously created with what we are currently discussing}". Similarly, P10 valued the feature for "\textit{making it easier to locate existing ideas when responding to other teammates}".

\paragraph{User Challenges and Feedback}
 
Participants identified several challenges they faced while using LADICA for co-located collaboration:

Firstly, several participants (P4, P6, P7) noted that while the ideation hints were helpful, at times the suggestions provided were slightly \textbf{off-topic or too general}. In addition, some participants (P2, P6, P9) felt that the growing number of automatically generated idea and relation hints could \textbf{clutter the whiteboard}, leading to elevated time and cognitive effort required to identify the insightful ones worthy of exploration and discussion.  

Participants also expressed concern that the reduced effort required to create idea notes might allow ``\textit{less diligent teammates to \textbf{appear engaged} while spending less time actively thinking}'' (P8). Furthermore, P14 thought that ``\textit{note-based ideation features could make people \textbf{focus more on refining their own ideas} instead of discussing with others}''. 

Lastly, some participants (P8, P10)  found the need to manually activate discussion-based information extraction and retrieval counter-intuitive, suggesting these features should display relevant information automatically in tune with the discussion flow.

The participants also offered suggestions to enhance the system's functionality. P8 proposed the introduction of features to facilitate direct comparisons between ideas, highlighting their advantages and disadvantages. The ability to summarize the ideas on the current page, as suggested by P2 and P5, would help teams understand the essence of shared content quickly, allowing more focused discussions. Similarly,  P9 recommended adding functionality to compare the results of analyzing ideas through different analytical lenses, which could enrich the depth of discussion. Lastly, P10 suggested recording individual edits in idea notes and employing LLMs to summarize the history of these edits, which could spark further discussion among contributors.

\section{Discussion}

\subsection{Generative AI for Supporting Human-Human Collaboration}

Human collaboration is a dynamic and interactive process where team members must process multiple cognitive activities simultaneously~\cite{colbry2014collaboration, fiore2018data}. This includes diverse top-down or bottom-up thought processes such as decomposing abstract team goals from a top-down perspective, generating a wide range of relevant ideas through divergent thinking, and employing analytical and convergent thinking to identify common themes in the group’s presented information~\cite{chiu2000group}. At the same time, it is crucial for team members to foster mutual awareness to spot collaboration opportunities during the teamwork process, maintain a working memory of ongoing conversations, and synchronize with the content in external shared tools~\cite{olson2000distance}. Our formative studies revealed that these cognitive challenges impede co-located teams from effectively contributing their ideas and engaging in frequent discussions during parallel work.

The design of LADICA demonstrates our human-centered approach that prioritizes the facilitation and augmentation essential cognitive tasks in human collaboration rather than taking over cognitive tasks for teams in designing LLM-based assistants. \zzrevision{Our three-layer conceptual framework (Figure \ref{fig:conceptual_framework}) echoed He et al.'s opinion \cite{he2024ai} that AI should play dynamic roles in facilitating human collaborations. Unlike many prior work that proactively recommend ideas for inspiration, LADICA preserves user initiatives to generating ideas, while providing scaffolding assistance for idea expansion, articulation and task structure decomposition. The participants found that these forms of assistance helped to bootstrap the initial ideation phase of the collaboration. However, they noted that with static prompting templates, the generative ideation support was sometimes off-topic, biased towards certain topics or overly generic. This suggests a need to incorporate real-time contextual group-generated information to enhance the relevance of the scaffolding prompts.}

\zzrevision{In addition to serving as an ideation stimulus, LADICA also leverages a generative AI model to support collective analytical and reflective thinking based on the shared workspace. Our study found that by facilitating idea grouping, connection, and retrieval during collaboration, the generative AI model helps teams build a shared mental model, align internal and external memory, and initiate cooperation. However, participants also expressed concerns that the potential over-reliance on AI support could hinder the natural discovery of collectively interesting topics through iterative discussion and occasionally reduce group engagement (P5, P9), particularly since the current system lacks detailed rationale or comparisons of elements within the same group to justify the suggested grouping. Besides, while participants appreciated the relational hints for highlighting discussion opportunities, some (P2, P6, P9) felt that the system often generated too many hints, which increased their mental load and distracted them during individual brainstorming. These challenges highlight the need for refined prompt strategies and design efforts to enhance user control and explainability in analytical processes and determine user preferences for the number of suggestions. }



\subsection{Human-AI Team in Co-Located Interaction}

Co-located interaction offers significant advantages for human teams, such as behavioral cues, immediate feedback, and the use of shared references in a common workspace~\cite{brignull2004introduction}. Nonetheless, integrating AI into these settings introduces distinct challenges not present in single-user or remote collaborations, as evidenced by our experiences with LADICA's design process. A notable challenge we encountered was the time-sensitive nature of co-located interaction. For designers, it is tempting to expedite the group discussion and decision-making process by automating most tasks for users, which can lead to risking the loss of critical human control and inclusiveness in this process. Additionally,  co-located teamwork features frequent informal discussions and loose collaboration forms. Therefore, imposing a rigid human-AI interaction workflow could undermine the valuable inherent flexibility in co-located settings. Lastly, when creating a shared user interface for co-located teams, we learned the importance of balancing between addressing individual needs with the group's collective needs in AI-supported tasks. \looseness=-1

LADICA's approach addresses these issues by leveraging AI to enhance, rather than replace, vital aspects of human collaboration, such as ideation, categorization, and open discussion. It avoids prescribing a fixed workflow, opting instead to support various forms of collaboration and thought processes that accommodate both individual and team needs.  For example, by facilitating the decomposition of the team goal, LADICA helps establish a common understanding early in the collaboration process.  It also uses affinity-based grouping and relational hints to promote teams to identify the relationships among their ideas. On an individual level, features like query and relation-based idea expansion encourage personal exploration and development of concepts.

Future research could explore ways to enrich AI's perception of the team's collaboration context through spatial interaction data from a shared display, thereby enhancing situational awareness through sensing technologies. Analyzing the spatial proximity of team members' focus points, for instance, could enable AI models to deduce the relevance of concurrent discussions or activities, offering advice that is more attuned to the dynamic interplay of team interactions.

\subsection{Interaction Modalities in Facilitating Co-Located Group Interactions}

Effective human-human collaboration in co-located settings relies on multi-modal communication channels, including shared workspace interactions, verbal exchanges, and behavioral observations~\cite{olson2000distance}. To align with these communication practices, LADICA incorporates support across both textual and auditory channels. It enhances on-screen collaboration by offering text-based suggestions for ideation, relationships, and grouping to stimulate team thinking and discussion. Simultaneously, LADICA supports verbal discussion in speech by facilitating the capture and summarization of key discussion points, thus supporting the team's cognitive processing across different communication mediums. Additionally, the system's relational hints prompt team members to engage in verbal discussions, further promoting interaction, inclusiveness, and collaborative synergy.

Beyond these modalities, there is potential to expand LADICA's functionality to include visual modalities in future developments. Employing Computer Vision (CV) technologies could allow AI-enhanced tools to interpret facial expressions, recognize gestures, and assess body postures, offering deep insights into team dynamics. Such capabilities could help in assessing the mood or level of engagement of team members, identifying moments where interventions might be beneficial, and determining the most appropriate forms of intervention. This direction not only broadens the scope of interaction modalities, but also enriches the understanding and facilitation of co-located team collaborations.

\section{Limitation \& Future Work}

The current version of LADICA has several technical limitations: First, the prompt templates for  various cognitive assistance are fixed and do not adapt to specific task scenarios or user context, which could lead to issues such as overly generic or irrelevant suggestions. Second, when multiple users simultaneously activate AI features, the system experiences a slowdown in response times from LLM, leading to  noticeable latency. Third, being web browser-based limits our system's interaction capabilities; some browsers do not fully support multi-touch functionality, restricting direct interaction with the shared display to a single user at a time. 

Regarding the study design, there are potential threats to the validity of our findings. For example, all of our participants were college students, which could introduce bias in the study results toward experiences based in an academic setting. It is plausible that co-located interaction in other contexts may have distinct needs and challenges. For instance, co-located collaboration in corporate settings may require consideration of the team's social hierarchy. Future work is also needed to further study the use of LADICA in broader categories of tasks beyond the trip planning and policy discussion scenarios used in our study. 

Moving forward, we plan to address these challenges and expand our research through several initiatives. We intend to deploy LADICA in actual classroom settings to observe its facilitation of real-world, co-located human collaboration. Efforts will also be made to enhance the AI features of LADICA, aiming to offer more dynamic cognitive support by predicting collaborators' relationships based on their physical proximity and interaction content. Additionally, we plan to explore the integration of multimodal interaction capabilities, including facial expression and body movement recognition, to improve usability and provide contexual cognitive assistance to a wider array of user groups.

\section{Conclusion}

This paper introduced LADICA, an AI-enhanced large shared display interface designed to support the meta- and macro-cognitive processes of ideation, grouping analysis, and discussion in co-located teams. Our formative studies identified several cognitive challenges in co-located team collaboration and gathered users' perspectives on AI intervention in team collaboration. In response, LADICA offered three layers of cognitive assistance to (1) encourage a dynamic group ideation process, (2) foster discussion and collaboration, (3) enable multi-faceted affinity-based analysis for group ideas, and (4) facilitate synchronization between the external display and verbal discussion. Our user studies showed that participants can successfully use LADICA for co-located collaboration and found the system useful in supporting their cognitive processes during collaboration.

\begin{acks}
This work was supported in part by NSF grants CNS-2426395, DRL-2437113, CCF-2211428, an NVIDIA Academic Hardware Grant, a Google Research Scholar Award, a Notre Dame-IBM Technology Ethics Lab Award, and a University of Notre Dame Strategic Framework Teaching Grant. Any opinions, findings or recommendations expressed here are those of the authors and do not necessarily reflect views of the sponsors. We would like to thank Xu Wang, Ron Metoyer and Diego Gómez-Zará for useful discussions.
\end{acks}

\bibliographystyle{ACM-Reference-Format}
\bibliography{main}

\newpage
\appendix
\newcolumntype{T}{>{\centering\arraybackslash}m{0.1\linewidth}}
\newcolumntype{Y}{>{\centering\arraybackslash}m{0.9\linewidth}}

\begin{table*}[]
\small
\centering
\section*{Appendix} 
\vspace{0.2cm}
\renewcommand{\arraystretch}{1.5}
\begin{tabularx}{\linewidth}{|T|Y|}
\cline{1-2}
  \textbf{Task} & \textbf{Prompt templates} \\
\cline{1-2}

\vspace{0.3cm}Relational hint generation & \vspace{0.3cm}\begin{itemize} [leftmargin=*, itemindent=0pt, nosep, align=left]
    \item Imagine you're the GPT-4 AI, assigned to support a team in their brainstorming sessions. During these sessions, every team member adds their notes to a whiteboard, each note touching on different facets a main subject or subtasks of a main goal. Your task involves analyzing a note that's currently being developed (source note). Your objective is to generate three notes that have a (selectWard) relationship with the source note.
    \item You are a very smart and experienced group work facilitator that is able to the relations between ideas on a shared whiteboard so as to stimulate team's mutual awareness and discussion. Your job is to identify the relations among ideas provided by team members. Please use the provided relation types. Plus, output the confidence score of prediction of each relation as well. Each pair of ideas can only have single relation type, do not repeatedly generate relations for a same pair. The relation list is as follows:
    \begin{itemize}
        \item "Is a": Indicates that one concept is a type or category of another.
        \item "Part of": Indicates that one concept is a part of another.
        \item "Used for": Describes what something is used for.
        \item "Capable of": Describes an action or activity that a concept is capable of doing.
        \item "At location": Indicates where something is typically found or where an event occurs.
        \item "Has a": Indicates that one concept possesses another.
        \item "Desires": Indicates a desire or need associated with a concept.
        \item "Causes": Describes an event or action that leads to a particular result.
        \item "Has property": Indicates a characteristic or property of a concept.
        \item "Synonym": Indicates that two concepts have the same or very similar meanings.
        \item "Antonym": Indicates that two concepts have opposite meanings.
        \item "Derived from": Indicates that one concept is derived from another, often used for words that have a common root or origin.
        \item "Instance of": Similar to IsA, but typically used for instances of a class or category.
    \end{itemize}
    \item Imagine you are the GPT-4 model, designed to assist a team in brainstorming sessions. Your task is to help them explore and understand the logical relationships between various text notes. Approach each note with an analytical mindset, drawing connections, identifying patterns, and suggesting possible links between different pieces of information. Encourage creativity and critical thinking, guiding the team through a constructive and collaborative brainstorming process. Your goal is to enhance their understanding and help them synthesize information in a meaningful way.
\end{itemize}  
\\
\cline{1-2}

\vspace{0.3cm}Affinity diagramming generation & \vspace{0.3cm}\begin{itemize} [leftmargin=*, itemindent=0pt, nosep, align=left]
    \item Imagine you are a very smart and experienced team leader that is able to identify the common interesting themes behind a group of ideas from different people. Your task is to identify the common underlying themes among ideas, and then group them based on your proposed themes. You need to propose different ways of grouping these items from diverse thinking perspectives. Plus, please also explain the brief rules of thumb of each way of grouping, as well as the short name of this grouping. Note that the user may provide some instructions as grouping direction, which we should follow if provided. Be creative and logical. Think of different ways of grouping first, then create the rules of thumb for each group, then create themes, then group ideas based on themes within each group. Note that an idea may be already assigned to a topic group ("pre\_topic", otherwise it is undefined). In this case, you need to include the old group name in the returned JSON as "pre\_topic" of the idea. The explanation of input JSON format is below. Do not use the same group topic as the original ones. Be creative and logical.
    \item For example, for ideas "Plan a trip to the Miami beach" (under group "trip schedule") and "Book flights from Chicago to Miami" (under group "travel"), the common themes could include "Cost", "Time", "Comfort".
\end{itemize} 
\\
\cline{1-2}

\end{tabularx}
\label{table:prompt_templates_1}
\end{table*}

\begin{table*}[]
\small
\centering

\renewcommand{\arraystretch}{1.5}
\begin{tabularx}{\linewidth}{|T|Y|}
\cline{1-2}

\vspace{0.3cm}Ideation and discussion hint generation & \vspace{0.3cm}\begin{itemize} [leftmargin=*, itemindent=0pt, nosep, align=left]
    \item Imagine you are the GPT-4 AI, integrated into a collaborative digital planning tool being used by a team to organize their spring break trip. Your role is to review the travel ideas and preferences already noted by the team members and generate discussion hints that facilitate a comprehensive and enjoyable planning process. These hints should encourage the team to explore various aspects of their trip, such as destination options, activities, budget considerations, accommodations, and any logistical requirements. Provide up to eight targeted discussion hints based on the topics, preferences, and constraints indicated by the team. Tailor your suggestions to ensure they are relevant and helpful, enhancing the team's ability to make informed decisions and create a memorable spring break experience. If the planning is still in its initial stages with few ideas noted, suggest starting points to guide the brainstorming.
    \newpage
    \item Imagine you're the GPT-4 AI, assigned to support a team in their brainstorming session. Your task is to inspire people on how to refine their brainstorming content. Make sure your suggestions are understandable and insightful.
    \item Imagine you're the GPT-4 AI, assigned to support a team in their brainstorming session. Your task is to improve the content based on the given suggestion. Make sure your improvement is understandable and insightful. The revision should be as short as possible.
    \item You are a smart team discussion facilitator that helps a group of people to collaborate on certain topics. During the discussion, different team members may create notes representing their ideas about the topic. Your task is to figure out the essential points between their ideas that they can discuss or collaborate on.
\end{itemize} 
\\
\cline{1-2}

\vspace{0.3cm}Speech recognition with information retrieval and extraction & \vspace{0.3cm}\begin{itemize} [leftmargin=*, itemindent=0pt, nosep, align=left]
    \item Imagine you are a very smart and careful agent that can help teams remember their generated content related to the current discussion. Your task is to identify the notes on the whiteboard relevant to the ongoing discussion. You are given a list of note contents and a transcript of the discussion. Return the relevant notes. For each note, also highlight the short segment of discussion relevant to the note. Return the results in the required list format.
    \item Imagine you are a very smart and careful team facilitator that can help teams to extract key information in their discussion. Your task is to identify those key pieces of information from a conversation transcript. You are given a list of idea note contents and a transcript of the discussion. Please return the list of key information, with each item being a short summary of unique key information that you think is relevant to the provided ideas. For each note, also highlight which existing note it may relate to. Return the results in the required JSON format.
\end{itemize} 
\\
\cline{1-2}

\end{tabularx}
\captionsetup{position=below}
\caption{Prompt templates used in Relational hint generation, Affinity diagramming generation, Ideation and discussion hint generation, and Speech recognition with information retrieval and extraction.}
\label{table:prompt_templates_2}
\end{table*}

\end{document}